%% file: resistiveMM.tex
\pdfoutput=1
\documentclass[preprint,12pt]{elsarticle}

\usepackage{amssymb}
\usepackage{amsmath}
\usepackage{float}
\usepackage{xcolor}

\usepackage{siunitx}
\usepackage{tikz}
\usepackage{lineno}
\usepackage{subcaption}
\usepackage{pdfpages}

\usepackage{hyperref}
\usepackage{comment}

\graphicspath{{figures/}}

\journal{Nuclear Instruments and Methods}

\begin{document}

\begin{frontmatter}





\title{Characterization of  resistive Micromegas detectors for the upgrade of the T2K Near Detector Time Projection Chambers}

%%%%%%%%%%%%%%%%%%%%%%%%%%%%%%%%%%%%
\input{author.tex}

%\cortext[cor]{Corresponding authors}
%%%%%%%%%%%%%%%%%%%

\begin{abstract}
The second phase of the T2K experiment is expected to start data taking in autumn 2022. An upgrade of the Near Detector (ND280) is under development and includes the construction of two new Time Projection Chambers called High-Angle TPC (HA-TPC). The two endplates of these TPCs will be paved with eight Micromegas type charge readout modules. The Micromegas detector charge amplification structure uses a resistive anode to spread the charges over several pads to improve the space point resolution. This innovative technique is combined with the bulk-Micromegas technology to compose the "Encapsulated Resistive Anode Micromegas" detector. A prototype has been designed, built and exposed to an electron beam at the DESY II test beam facility.\\
The data have been used to characterize the charge spreading and to produce a RC map. Spatial resolution better than 600 $\text{\textmu}\textrm{m}$  and energy resolution better than 9\% are obtained for all incident angles.  These performances fulfil the requirements for the upgrade of the ND280 TPC.
%%%%%%%%%%%%
\end{abstract}

\begin{keyword}
Resistive Micromegas, T2K Near Detector Time Projection Chambers



\end{keyword}

\end{frontmatter}

\tableofcontents
\newpage
% \linenumbers
\section{Introduction}
\input{intro.tex}
\label{sec:introduction}
%%%%%%%%%%%%%%%%%%%%%%%%%%%%%%%%%%%%%%%%%%%%%%%%%%%%%%%%%%%%%%%%%
\section{Resistive Micromegas}
\label{sec:Micromegas}
\input{eram.tex}

%%%%%%%%%%%%%%%%%%%%%%%%%%%%%%%%%%%%%%%%%%%%%%%%%%%%%%%%%%%%%%%%%
\section{Experimental setup}
\label{sec:setup}
\input{setup.tex}
%%%%%%%%%%%%%%%%%%%%%%%%%%%%%%%%%%%%%%%%%%%%%%%%%%%%%%%%%%%%%%%%%
\section{Collected data}
\label{sec:data}
\input{cOllectedData.tex}

%%%%%%%%%%%%%%%%%%%%%%%%%%%%%%%%%%%%%%%%%%%%%%%%%%%%%%%%%%%%%%%%%
\section{ Charge spreading characterisation}
\label{sec:spread}
\input{spread.tex}
%%%%%%%%%%%%%%%%%%%%%%%%%%%%%%%%%%%%%%%%%%%%%%%%%%%%%%%%%%%%%%%%%
\section{Track reconstruction}
\label{sec:reco}
\input{reco.tex}

%%%%%%%%%%%%%%%%%%%%%%%%%%%%%%%%%%%%%%%%%%%%%%%%%%%%%%%%%%%%%%%%%%%%%%%%%%%%
\section{Spatial resolution}
\label{sec:spatial}
\input{spatial.tex}

%%%%%%%%%%%%%%%%%%%%%%%%%%%%%%%%%%%%%%%%%%%%%%%%%%%%%%%%%%%%%%%%%
\section{Deposited energy resolution}
\label{sec:dedx}

\input{dedx2.tex}
%%%%%%%%%%%%%%%%%%%%%%%%%%%%%%%%%%%%%%%%%%%%%%%%%%%%%%%%%%%%%%%%%
\section{RC map calculation}
\label{sec:rcmap}
\input{rcmap.tex}
%%%%%%%%%%%%%%%%%%%%%%%%%%%%%%%%%%%%%%%%%%%%%%%%%%%%%%%%%%%%%%%%%
\section{Conclusions}
\label{sec:conclusion}
\input{conclusions.tex}
%%%%%%%%%%%%%%%%%%%%%%%%%%%%%%%%%%%%%%%%%%%%%%%%%%%%%%%%%%%%%%%%%
\section*{Acknowledgements}
The measurements leading to these results have been performed at the Test Beam Facility at DESY Hamburg (Germany), a member of the Helmholtz Association.
The authors would like to thank the technical team at the DESY II accelerator and test beam facility for the smooth operation of the test beam and the support during the test beam campaign. \\
We acknowledge the support of CEA and CNRS/IN2P3, France; DFG, Germany; INFN, Italy; National Science Centre (NCN) and Ministry of Science and Higher Education (Grant No. DIR/WK/2017/05), Poland; MINECO and ERDF funds, Spain. \\
In addition, participation of individual researchers and institutions has been further supported by H2020 Grant No. RISE-GA822070-JENNIFER2 2020, MSCA-COFUND-2016 No.754496, ANR-19-CE31-0001, RFBR grants \#19-32-90100, the Ministry of Science and Higher Education of Russia (contract \#075-15-2020-778) the Spanish Ministerio de Econom\'{i}a y Competitividad  (SEIDI - MINECO) under Grants No.~PID2019-107564GB-I00 and SEV-2016-0588. IFAE is partially funded by the CERCA program of the Generalitat de Catalunya.
%%%%%%%%%%%%%%%%%%%%%%%%%%%%%%%%%%%%%%%%%%%%%%%%%%%%%%%%%%%%%%%%%




%\section*{References}

\bibliographystyle{elsarticle-num}
\bibliography{bibliography}

\end{document}

%% file: author.tex
\author[saclay]{D.~Atti\'e}
\author[ifj]{M.~Batkiewicz-Kwasniak}
\author[lpnhe]{P.~Billoir}
\author[lpnhe]{A.~Blanchet}
\author[lpnhe]{A.~Blondel}
\author[saclay]{S.~Bolognesi}
\author[saclay]{D.~Calvet}
\author[bari]{M.G.~Catanesi}
\author[legnaro]{M.~Cicerchia}
\author[padova]{G.~Cogo}
\author[saclay]{P.~Colas}
\author[padova]{G.~Collazuol}
\author[saclay]{A.~Delbart}
\author[lpnhe]{J.~Dumarchez}
\author[saclay]{S.~Emery-Schrenk}
\author[padova]{M.~Feltre}
\author[lpnhe]{C.~Giganti}
\author[legnaro]{F.~Gramegna}
\author[padova]{M.~Grassi}
\author[lpnhe]{M.~Guigue}
\author[aachen]{P.~Hamacher-Baumann}
\author[saclay]{S.~Hassani\fnref{fnref1}}
\author[padova]{F.~Iacob}
\author[ifae]{C.~Jes\'{u}s-Valls}
\author[wut]{R.~Kurjata}
\author[padova]{M.~Lamoureux}
\author[saclay]{M.~Lehuraux}
\author[padova]{A.~Longhin}
\author[ifae]{T.~Lux}
\author[bari]{L.~Magaletti}
\author[legnaro]{T.~Marchi}
\author[saclay]{A.~Maurel}
\author[lpnhe]{L.~Mellet}
\author[padova]{M.~Mezzetto}
\author[saclay]{L.~Munteanu}
\author[lpnhe]{Q.~V.~Nguyen}
\author[lpnhe]{Y.~Orain}
\author[padova]{M.~Pari}
\author[lpnhe]{J.-M.~Parraud}
\author[bari]{C.~Pastore}
\author[padova]{A.~Pepato}
\author[lpnhe]{E.~Pierre}
\author[lpnhe]{B.~Popov}
\author[ifj]{H.~Przybiliski}
\author[aachen]{T.~Radermacher}
\author[bari]{E.~Radicioni}
\author[saclay]{M.~Riallot}
\author[aachen]{S.~Roth}
\author[wut]{A.~Rychter}
\author[padova]{L.~Scomparin}
\author[aachen]{J.~Steinmann}
\author[lpnhe,inr]{S.~Suvorov\fnref{fnref2}}
\author[ifj]{J.~Swierblewski}
\author[lpnhe]{D.~Terront}
\author[aachen]{N.~Thamm}
\author[lpnhe]{F.~Toussenel}
\author[bari]{V. Valentino}
\author[saclay]{G.~Vasseur}
\author[lpnhe]{U.~Yevarouskaya}
\author[wut]{M.~Ziembicki}
\author[lpnhe]{M.~Zito}
%
%%%%%%%%%%%%%%%%%%%%%%%%%%%%%%%%%%%%%%%%%%%%%%%%%%%%%%%%%%%%%%%%%%%%%%%%%%%%%%%%%%%%
\address[saclay]{IRFU, CEA, Universit\'e Paris-Saclay, Gif-sur-Yvette, France}
\address[ifj]{H. Niewodniczanski Institute of Nuclear Physics PAN, Cracow, Poland}
\address[lpnhe]{LPNHE, Sorbonne Universit\'e, Universit\'e de Paris, CNRS/IN2P3, Paris; France}
\address[bari]{INFN sezione di Bari, Universit\`a di Bari  e Politecnico di Bari, Italy}
\address[legnaro]{INFN: Laboratori Nazionali di Legnaro (LNL), Padova , Italy}
\address[padova]{INFN Sezione di Padova and Universit\`a di Padova, Dipartimento di Fisica e Astronomia, Padova, Italy}
\address[aachen]{RWTH Aachen University, III.~Physikalisches Institut, Aachen, Germany}
\address[ifae]{Institut de Fisica d'Altes Energies (IFAE), The Barcelona Institute of Science and Technology, Bellate Spain}
\address[wut]{Warsaw University of technology, Warsaw, Poland}
\address[inr]{Institute for Nuclear Research of the Russian Academy of Sciences, Moscow, Russia}
%%%%%%%%%%%%%%%%%%%%%%%%%%%%%%%%%%%%%%%%%%%%%%%%%%%%%%%%%%%%%%%%%%%%%%%%%%
\cortext[cor1]{Corresponding author}
\fntext[fnref1]{samira.hassani@cea.fr}
\fntext[fnref2]{sergey.suvorov@lpnhe.in2p3.fr}

%% file: intro.tex
% !TEX root = resistiveMM.tex
%text written by Samira inspired from Sara's contribution to the PRR document
T2K~\cite{Abe:2011ks} is a long-baseline neutrino oscillation experiment exploiting a muon neutrino beam produced by the JPARC accelerator complex in Japan. The T2K experiment  includes the secondary neutrino beamline, a set of near detectors (INGRID and ND280) and the far detector SuperKamiokande.

%The phase 2 of T2K is expected to start data taking in autumn 2022. An upgrade of the Near Detector (ND280) is under development and includes the construction of two new Time Projection Chambers called High-Angle TPC (HA-TPC). The two endplates of these TPCs will be paved with eight Micromegas type charge readout modules. The Micromegas detector charge amplification structure uses a resistive anode to spread the charges over several pads to improve the space point resolution. This innovative technique is combined with the bulk-Micromegas technology ~\cite{Abgrall:2010hi} to compose the "Encapsulated Resistive Anode Micromegas" (ERAM) detector. This paper describes the performance of ERAM module.\\

T2K provided the first evidence of non-zero mixing angle $\theta_{13}$~\cite{Abe:2011sj} and discovered the appearance of electron neutrinos in a muon neutrino beam~\cite{Abe:2011sj, Abe:2013xua, Abe:2013hdq}. Combining T2K data with precise $\theta_{13}$ measurement from reactor experiments, T2K has recently reported hints of large Charge-Parity (CP) violation in the leptonic sector~\cite{Abe:2019vii}, excluding CP-conservation at about the 2$\sigma$ Confidence Level. 

%T2K discovered the appearance of electron neutrinos in a muon neutrino beam~\cite{Abe:2011sj, Abe:2013xua,Abe:2013hdq}, providing the first evidence for non-zero mixing angle $\theta_{13}$. 
%Combining T2K data with precise $\theta_{13}$ measurement from neutrino experiments at reactors, T2K has provided for the first time limits at 3$\sigma$ confidence level on some of the values of the phase $\delta_{CP}$~\cite{Abe:2019vii}, providing hints of maximally-violated Charge-Parity symmetry in neutrino oscillation (the case of CP-conservation being excluded at about 2$\sigma$). 

The T2K collaboration is now preparing for the second phase of the experiment (T2K-II), starting in Fall 2022, which will exploit the upgrade of the beam from 500~kW to 750~kW. T2K-II will collect in total more than $10^{22}$ Protons-On-Target (POT), including the $3.6 \times 10^{21}$~POT already collected, thus enabling 3$\sigma$ sensitivity on CP-Violation, in case of maximally violated CP, as currently indicated by the T2K results~\cite{Abe:2019vii}. In order to cope with such increased statistics, improved control of the relevant systematic uncertainties is needed. To this aim, an upgrade to ND280 is being constructed.

ND280 is a magnetized multi-purpose detector that measures, before the oscillation, the neutrino differential rate versus the kinematics variables of the particles, mostly muons, that result from the neutrino interaction with nuclei in dense parts of the detector. This allows one to constrain the neutrino flux and the neutrino-nucleus interaction cross-section. 
The present ND280 consists of two main parts: an upstream $\pi^0$ detector (P0D) and a downstream tracker which includes two Fine Grained Scintillators (FGD) interleaved with three vertical Time Projection Chambers (TPCs). The P0D and the tracker are surrounded by an electromagnetic calorimeter (ECAL) and by the UA1 magnet providing a 0.2~T magnetic field. The magnet yoke is further instrumented with a Side Muon Range Detector (SMRD).

The role of ND280 in the T2K oscillation analysis is crucial, allowing one to constrain the uncertainty on the expected number of events at the far detector to 4-5\%. 
ND280 measurements are performed on different targets (Carbon and Oxygen) and rely on the precise measurement of the muon momentum measured by the TPCs, with a momentum resolution of 10\% at 1~GeV~\cite{Abgrall:2010hi}. Better resolution is not needed since the determination of the incoming neutrino energy from the outgoing muon momentum in $\nu_{\mu}$ charged-current interactions is limited by the uncertainty induced by the Fermi motion of the nucleon in the nucleus.

Another important result of ND280 is the measurement of the $\nu_e$ contamination in the beam~\cite{Abe:2014usb,Abe:2020vot}, that constitutes one of the two main backgrounds to the $\nu_e$ appearance in the electron-like sample selected at Super-Kamiokande. This analysis, based on the Particle Identification (PID) capabilities of the TPCs and of ECAL, is possible thanks to the deposited energy resolution of 8\% in the TPCs that allows sufficient e-$\mu$ separation between a few hundred MeV and $\sim$2 GeV.

%ND280 is a magnetized multi-purpose detector able to measure the wrong-sign contamination in the beam (neutrino background in the antineutrino-beam mode and viceversa) and to measure the rate of neutrino interactions in different targets (Carbon, Water) as a function of the neutrino energy. In order to be able to reconstruct, as much precisely as possible, the neutrino energy, ND280 needs to measure and identify the particles produced by neutrino interactions in the detector with good momentum resolution, low momentum threshold and large angular acceptance.

%The present ND280 consists of two main parts: an upstream $\pi^0$ detector (P0D) and a downstream tracker including two Fine Grained Scintillators (FGD) interleaved with three vertical Time Projection Chambers (TPCs). The P0D and the tracker are surrounded by an electromagnetic calorimeter (ECAL) and by the UA1 magnet providing a 0.2~T magnetic field. The magnet yoke is further instrumented with a Side Muon Range Detector (SMRD).
The upgrade of ND280 consists in substituting the P0D with a new tracker, similar to the existent one but with a horizontal orientation parallel to the neutrino beam, and able to detect particles transverse to the beam, for which the acceptance of the current tracker is small. The new tracker includes a 3-dimensional scintillator target for neutrino interactions(Super-FGD) made of about 2 million 1~$\text{cm}{^3}$~cubes, read out by wavelength shifting fibers in the 3 directions~\cite{Blondel_2020}.  On the top and the bottom of the Super-FGD, two High-Angle TPCs (HA-TPCs) will be installed. The new tracker system will be surrounded by six Time-of-Flight modules.
This new detector configuration will allow one to improve the angular acceptance of ND280, being close to the full 4$\pi$ phase-space accessible at SuperKamiokande. In addition, the better tracking performances of the super-FGD will allow to improve the reconstruction of the hadronic part of the neutrino interactions, that will be exploited in combination with the muon kinematics. Also, more target mass for neutrino interactions yields more data statistics. 

Each endplate of the new TPCs will be instrumented with eight Micromegas type charge readout modules. The Micromegas detector charge amplification structure uses a resistive anode to "spread" the charges over several pads to improve the space point resolution: a signal is induced in pads adjacent to the pad receiving directly electrons from the avalanche, and the information from those additional signals allows one to better measure the track position. This effect is different from the charge sharing between pads resulting from the diffusion of the primary electrons in the TPC gas. This innovative technique is combined with the bulk-Micromegas technology ~\cite{Abgrall:2010hi} to compose the "Encapsulated Resistive Anode Micromegas" (ERAM) detector. Performances of a prototype of an ERAM detector exposed to a test beam at CERN were shown in~\cite{Attie:2019hua}.
This paper describes the performance of one prototype ERAM module, built with the same design that will be used for the HA-TPCs, exposed to an electron beam at DESY. With respect to our previous studies reported in~\cite{Attie:2019hua}, in this paper we characterize both  spatial and dE/dx resolution as a function of the angle of the track projection reconstructed in the ERAM plane, with the side of the ERAM plane, corresponding to the sides of the ERAM pads. 

%In view of the increased statistics, an important limiting factor in the measurement of neutrino oscillations at T2K are the control of systematic uncertainties on neutrino-nucleus interactions, notably in the extrapolation of near detector measurements to predictions at far detector. In order to minimize such uncertainties ND280 acceptance should cover, as much as possible, the full $4\pi$ phase-space accessible at SuperKamiokande. The 3D structure of the Super-FGD and the horizontal geometry of High-Angle TPCs will provide new acceptance coverage for high-angle tracks. 

%For the measurement of electron neutrino appearance (and thus $\delta_{CP}$) the most relevant background is the rate of electron neutrinos produced in the beam itself by pion and kaon decays. To control such background and to minimize the uncertainty of electron-neutrino over electron-antineutrino cross-section, which affects directly the measurement of $\delta_{CP}$, the measurement of electron neutrino rate at ND280 is important. This requires good particle identification (PID) performances for muon versus electrons in the TPCs, imposing a constrain on the dE/dx resolution of the order of 8\% on minimum ionizing particles in order to provide a sufficient e-$\mu$ separation. This dE/dx resolution will also enable good separation between minimum ionizing particles and protons, up to momenta of about 1 GeV, which cover most of the tracks at T2K energy.

%The already installed vertical TPCs feature a momentum resolution of the order of 10\% at 1~GeV~\cite{Abgrall:2010hi}. 
The HA-TPC should match the performance of existing TPCs in terms of momentum and dE/dx resolution. In addition, the super-FGD will enable the reconstruction of low momentum protons and neutrons~\cite{Abe:2019whr}, thus allowing one to reconstruct the incoming neutrino energy more precisely and {\it effectively} correct for the Fermi motion: an improved track momentum resolution in the HA-TPC, even beyond the previous specifications, is therefore useful to improve the neutrino energy resolution and, ultimately, the precision on neutrino oscillation measurements.

%% file: eram.tex
% !TEX root = resistiveMM.tex
%%Text written by Samira inspired from PRR document
The ERAM detector uses the bulk-Micromegas technology invented in 2004 by a collaboration between the CERN/EP-DT-EF PCB workshop and CEA-IRFU~\cite{Giomataris:2004aa}. 
It was developed and used for the construction of the 72 bulk-Micromegas modules (9 $\textrm{m}^{2}$ total area) which equip the three T2K/ND280 vertical TPCs. 
%trying to create a proposal: The woven micromesh of the Micromegas detector is obtained by sandwiching it between two layers of the same insulating material, pyralux, and then by exposing it to UV radiation allowing to keep only some precise positions. 
The woven micromesh of the Micromegas detector is sandwiched between two layers of the same insulating material (pyralux) and exposed to UV radiation at the location where the pyralux is kept on top of the pad-segmented anode PCB after chemical development.

\begin{figure}[htbp]
\centering
\includegraphics[width=.95\textwidth,origin=c,angle=0]{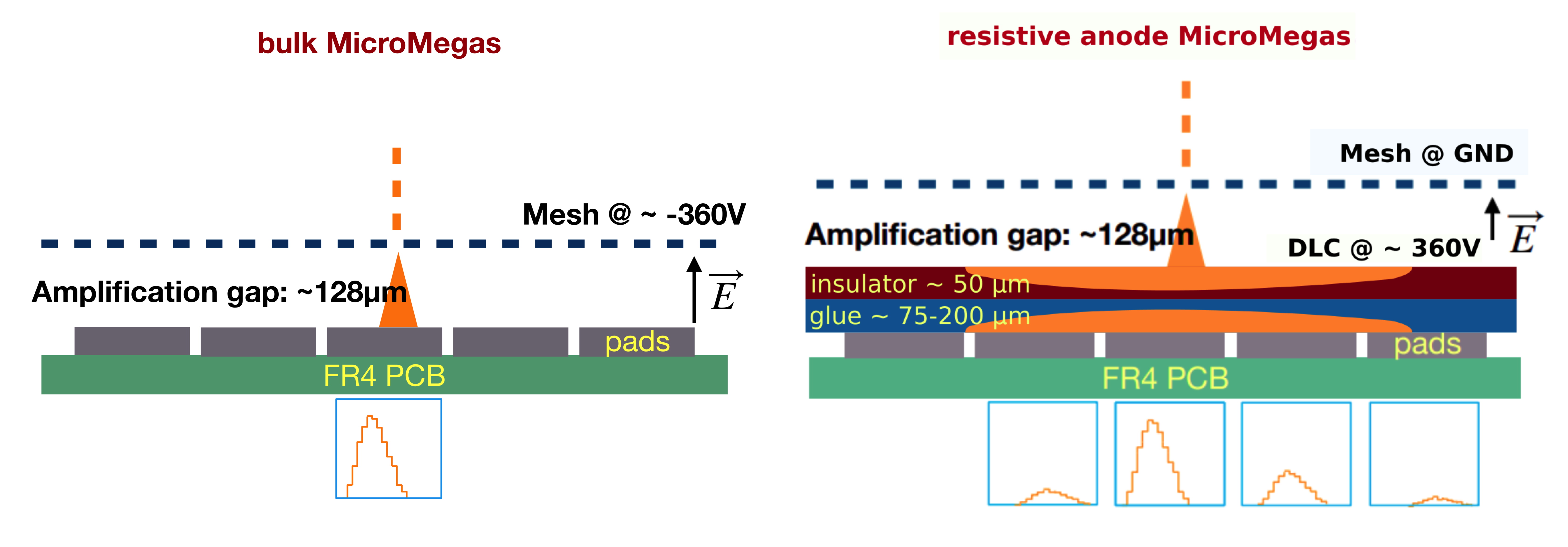}
\caption{\label{fig:MM_sketches}  A sketch of the Standard Bulk Micromegas layout (left) and the Resistive anode Micromegas layout (right).}
\end{figure}

The ERAM detector is a 128 $\text{\textmu}\textrm{m}$ amplification gap bulk-Micromegas using the standard SD45/18 304L woven micromesh built on top of a resistive anode PCB. 
When a track crosses the gas volume of a TPC, it generates a cloud of ionized electrons. These electrons are drifted to the anode readout plane of the TPC under a uniform electric field. On the readout plane, an avalanche is generated by a high electric field in the Micromegas amplification region, called DLC voltage in the following of this paper. The 3D position of the track is then reconstructed from the arrival time and the position of the avalanche on the readout plane.
%in order to be multiplied by avalanche in a high electric field generated by a MWPC or MPGD. Their time of arrival is measured and they are localized in the anode readout plane to derive the 3D ionizing track. 
In the case of the bulk-Micromegas, the electron avalanche in the amplification gap is quite narrow with respect to the pad size and therefore the position resolution is often limited by the pad size. The resistive anode Micromegas, first introduced and extensively studied by the ILC-TPC collaboration ~\cite{Dixit_DLC_nim2004}, provides a way to induce a signal on a larger number of pads allowing for a better reconstruction of the track position. A sketch of the bulk and the resistive Micromegas concepts are presented in~\autoref{fig:MM_sketches}. 

The anode, segmented in pads, is covered by a foil of insulating material, which has a thin resistive layer on its top side. 
The ERAM detector uses a Diamond-Like Carbon (DLC) thin layer sputtered on a 50 ${\rm\text{\textmu} m}$ thick APICAL (Kapton) insulator sheet.
The avalanche is then naturally quenched because the potential difference locally drops in presence of a high charge density. 
The resistive layer acts like a 2-D RC network and the charge deposited by the avalanche induces a signal also on the adjacent pads, looking as if the charge spreads naturally with time following a Gaussian behaviour. 
For a point charge deposited at $r=0$ and $t=0$, the charge density as a function of radius $r$ and time $t$ is given by:
\begin{equation}
\label{eq:RMM}
\rho(r,t) = \frac{RC}{4\pi t} e^{\frac{-r^2 RC}{4t}}  
%\frac {RC} {2 t} e^{- r^2 RC / (4 t)}
\end{equation}
where $R$ is the resistivity per unit area and $C$ is the capacitance per unit area. For this structure, the capacitance $C$ is defined by the distance between the resistive layer and the grounded pads (glue thickness plus APICAL foil). 
The insulating layer thickness determines the width of the induced charge spread $\sigma_{t}$ at a given time. When measured, due to integration of the charge collected by a front-end electronics of shaping time $t$, this spread is of the order of $\sigma_{t}=\sqrt\frac{2t}{RC}$. 

The resistive anode provides mainly two advantages: by spreading the charge between neighbouring pads, it improves greatly the resolution with respect to the pitch$/\sqrt{12}$ provided by a mere hodoscope, and it suppresses the formation of sparks and limits their intensity. 
A further and novel improvement of this technique is a new High Voltage powering scheme, where the mesh is set to ground and the anode to a positive amplification voltage. 
%before: A further and novel improvement of this technique is a new High Voltage powering scheme, where the mesh is set to ground while the anode is set to a positive amplification voltage. 
The insulation of the resistive anode from the pads, hence from the electronics, ensures a safe operation by a capacitive coupling readout and thus allows us to remove the cumbersome anti-spark protection circuitry of electronic read-out boards necessary in the case of the standard bulk readout.
%before: The insulation of the resistive anode from the pads, hence from the electronics, allows a safe operation by a capacitive coupling readout, thus allowing to get rid of the cumbersome anti-spark protection circuitry necessary in the case of the standard bulk readout.

At the end of 2017, the first series of prototypes were produced and tested to assess the feasibility of large charge spreading by a low resistivity anode. 
The vertical TPCs PCB, with an active area of 36$\times$34~cm$^2$ covered by 0.97$\times$0.69~cm$^2$ pads, was adapted to build an ERAM structure with "on-shelf" 2.5 MOhm/$\Box$ DLC foils. 
This first prototype was tested with a particle beam at CERN in 2018 and its performances are summarized in~\cite{Attie:2019hua}.

In fall 2018, the global design of the ND280 upgrade detector was fixed. 
The sub-detector envelops were defined and the size of the ERAM modules fixed to be $420 \times 340$ mm$^2$ with $32\times36$ rectangular pads of size $10.09 \times 11.18$~mm$^2$. 
The ERAM module studied in this paper has a resistivity close to the required one of 200 kOhm/$\Box$ using DLC foils stack on a 75 ${\rm\text{\textmu} m}$ glue layer.
The resistivity maps measured at critical steps of the manufacturing of the DLC foil shows a non uniformity of 20\%. 
The detector is readout with two analogue 576 channels AFTER based Front-End Cards (eight 72 channels AFTER ASIC per card). The AFTER chip, already used for the existing TPCs of T2K and for the test beam described in~\cite{Attie:2019hua} allows tuning several parameters such as the gain, the shaping time and the sampling time.
This prototype was tested in a test beam at DESY in June 2019.

%% file: setup.tex
% !TEX root = resistiveMM.tex

%%text witten by Samira with the help of David
%%%%%%%%%%%%%%%%%%%%%%%%%%%%%%
The prototype has been exposed to an electron beam at the DESY II test beam facility~\cite{Diener:2018qap}. 
DESY II provides electron beams of $1-6~\rm {GeV/c}$ at a rate of up to several kHz, depending on the chosen momentum. 
%previously: from  $1-6~\rm {GeV/c}$ at a rate of up to several kHz, depending on the chosen momenta. 
In the test beam area TB24/1, a large-bore superconducting solenoid, called PCMAG, provides a magnetic field of up to $1.25 ~\rm{T}$. 
%But a field of $0.2 ~\rm{T}$ will be used, as in the T2K experiment.
%Previously: In the test beam area TB24/1, a large-bore superconducting solenoid is installed, called PCMAG which can provide a magnetic field of up to $1.25 ~\rm{tesla}$. 
The magnet is mounted on a movable platform, which allows the setup to be moved horizontally and vertically, perpendicular to the beam line, as well as rotated by $\pm \SI{45}{\degree}$ in the horizontal plane. The platform can position the device under test with a precision of about \SI{0.2}{\mm} horizontally, \SI{0.1}{\mm} vertically, and within \SI{0.1}{\degree} in angle. 
%Previously: The magnet is mounted on a movable stage, which allows the setup to be moved horizontally and vertically, perpendicular to the beam line, as well as rotated by $\pm \SI{45}{\degree}$ in the horizontal plane. The stage can position the device under test with a precision of about \SI{0.2}{\mm} horizontally, \SI{0.1}{\mm} vertically, and within \SI{0.1}{\degree} in angle. 

\begin{figure}[!ht]
    \centering
    \begin{minipage}{0.49\linewidth}
        \centering
        \includegraphics[width=\linewidth]{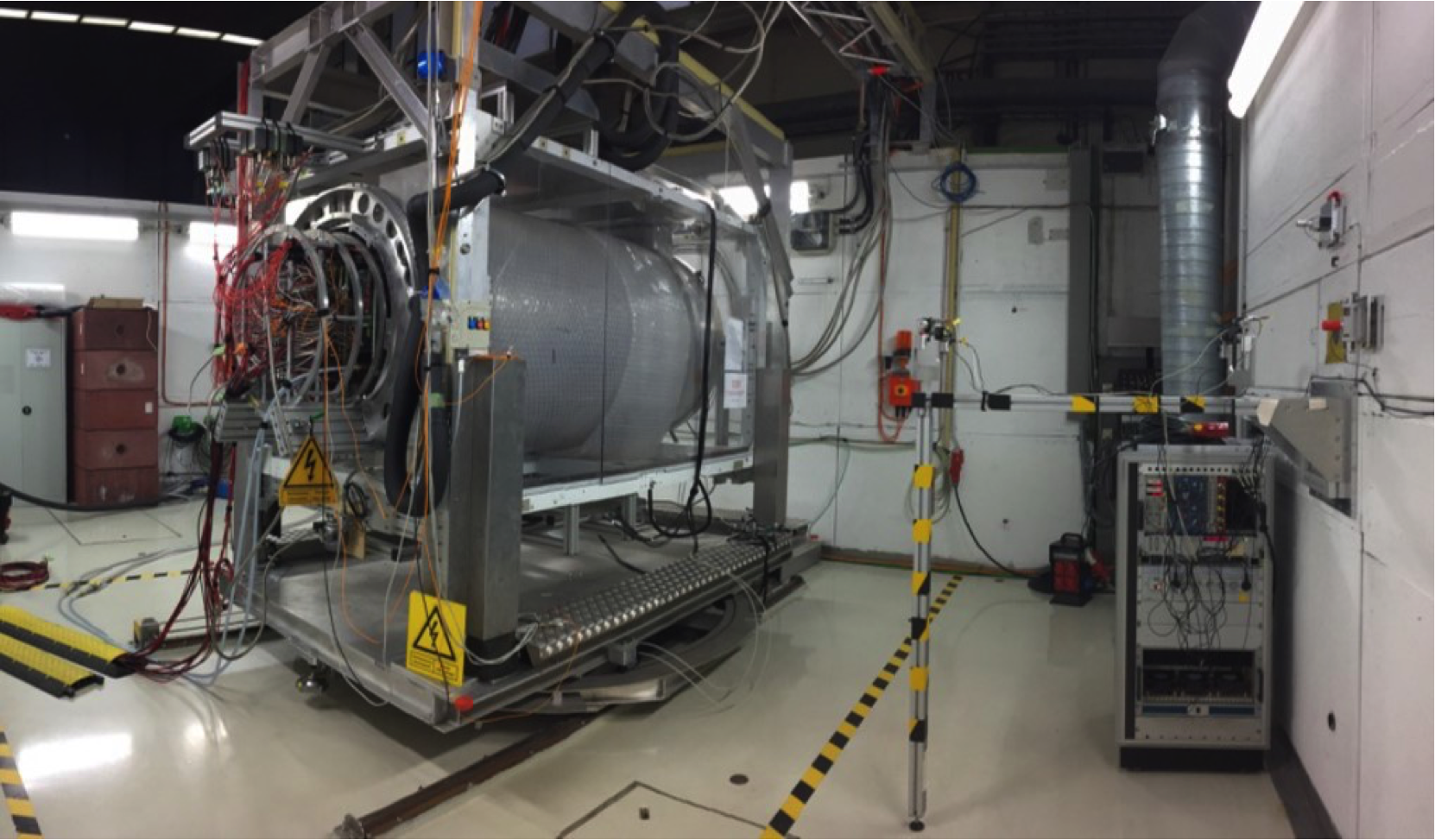} 
    \end{minipage}
    \begin{minipage}{0.49\linewidth}
        \centering
        \includegraphics[width=\linewidth]{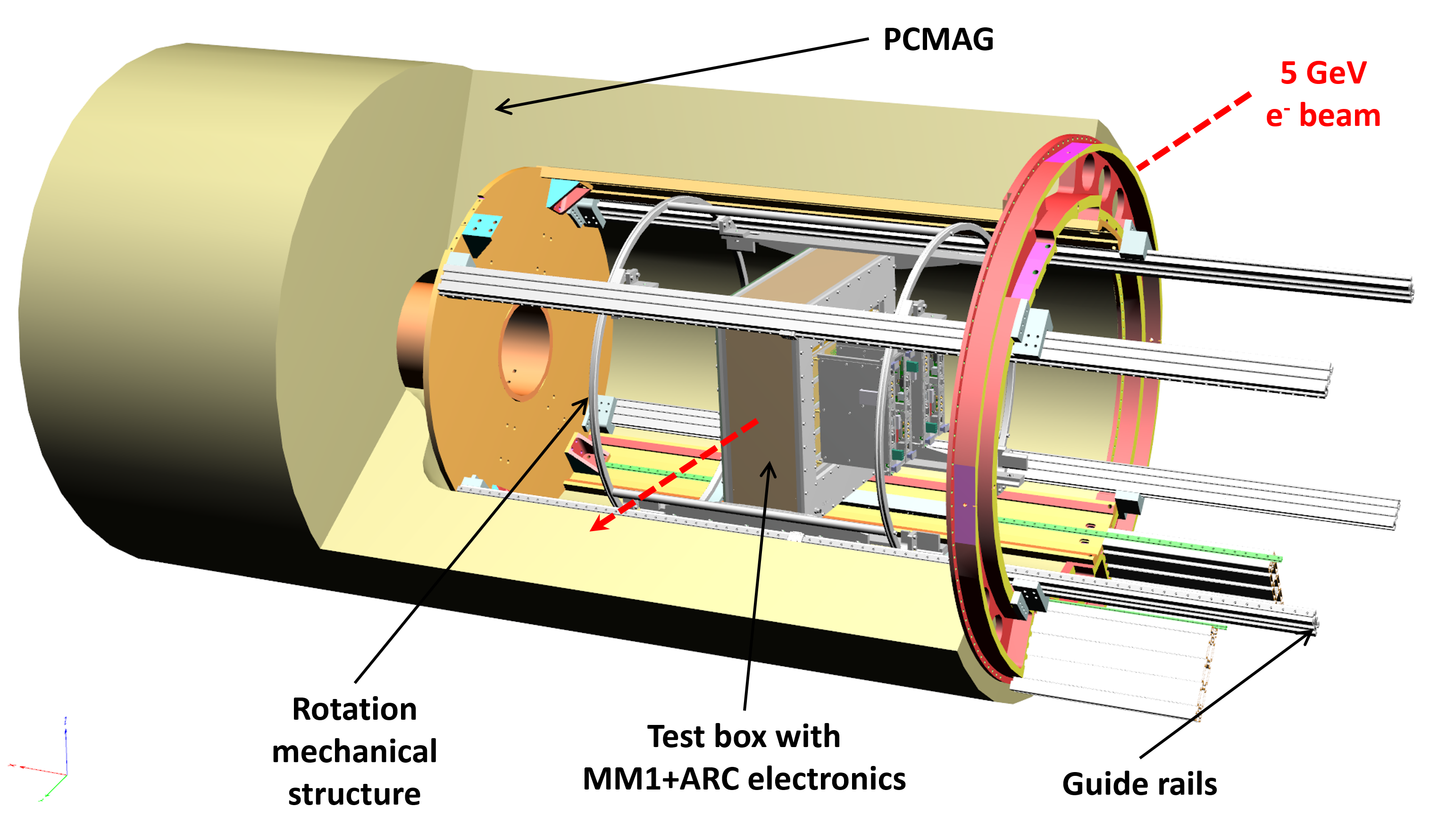} 
    \end{minipage}
    \caption{View of the setup.}
    \label{fig:DESY_setup}
\end{figure}

Inside the bore of the magnet, a rail system is installed on which test devices can be mounted at different positions within the magnet. The TPC prototype is supported on a sled, which can move in and out of the magnet and can be used to rotate the chamber around the magnetic field axis as illustrated in~\autoref{fig:DESY_setup}. 

Usually, the magnet is positioned perpendicular to the beam, so the magnetic field is also perpendicular to the beam, in a horizontal plane. The ERAM module is a vertical plane parallel to the beam, but perpendicular to the magnetic field. The walls of the magnet present about \SI{20}{\percent} of a radiation length, so that an electron beam of $6~\rm {GeV/c}$ easily penetrates the magnet and the device under test. A set of four consecutive scintillation counters, of which each has an area of approximately \SI{2.5 x 2.5}{\cm}, is mounted about \SI{1.5}{\m} in front of the magnet. The coincidence between them is used as a beam trigger. In addition, a second set of scintillation counters above and below the magnet provides a cosmic trigger for tests without beam. 

The prototype has been placed
inside the magnetic field of 0.2 T provided by the PCMAG magnet and has been operated for these measurements with a gas mixture of 95\% argon, 3\% tetrafluoromethane (CF$_4$), and 2\% isobutane (iC$_4$H$_{10}$) that is the same gas mixture as the one used in the existing TPCs of ND280. The gas quality was constantly monitored during the measurement. For the results reported in this paper, the oxygen contamination was around $30~\rm{ppm}$ at a gas flow rate of $30~\ell/h$. The chamber was operated at atmospheric pressure. Ambient temperature and pressure were constantly monitored.

%% file: collectedData.tex
%editor : Samira 
 The collected data is used to analyze in detail the role of all the relevant parameters (electronics peaking time, DLC voltage, drift distance) to fully characterize the charge spreading, the resistive foil uniformity, and to ensure a performance satisfying the ND280 upgrade requirements.
 %Previously: it had the main goal to fully characterize the charge spreading, the resistive foil uniformity and to ensure a performance satisfying the ND280 upgrade requirements.
The tests at DESY were done in a short chamber with a 15 cm drift distance. The prototype was operated at a voltage of 360 V. The settings chosen for the AFTER chip were a sampling time of 40 ns, a peaking time of 412 ns or 200 ns, and a gain of 120 fC. 
%Previously: The prototype was operated with a voltage of 360 V. The settings chosen for the AFTER chip were a sampling time of 40 ns, a peaking time of 412 ns or 200 ns and a gain of 120 fC. 
 
 The results presented in this paper were obtained with electrons with momenta varying from 0.5 to 5 GeV/c. We have carried out drift distance scans with seven points, spaced by 2 cm, at $B=0~\textrm{T}$ and $B=0.2~\textrm{T}$, peaking time of 200 ns or 412 ns, and high voltage of 370 V and 360 V. In addition a scan of the DLC voltage, varied from 330 V to 400 V, was performed.
 
\begin{figure}[!ht]
    \centering
    \begin{minipage}{0.45\linewidth}
        \centering
        \includegraphics[width=\linewidth]{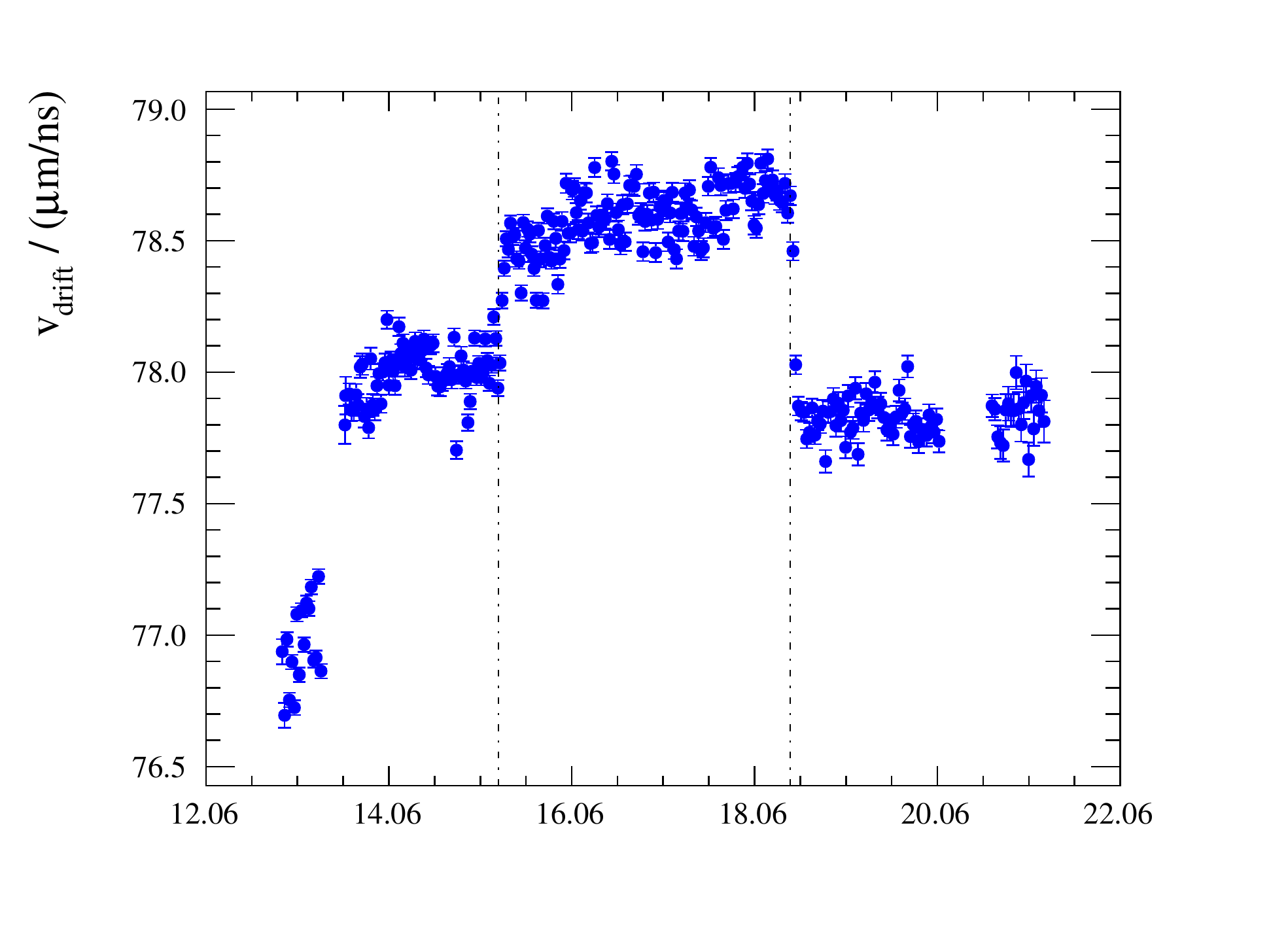}\\a
    \end{minipage}%
    ~%
    \begin{minipage}{0.45\linewidth}%
        \centering
        \includegraphics[width=\linewidth]{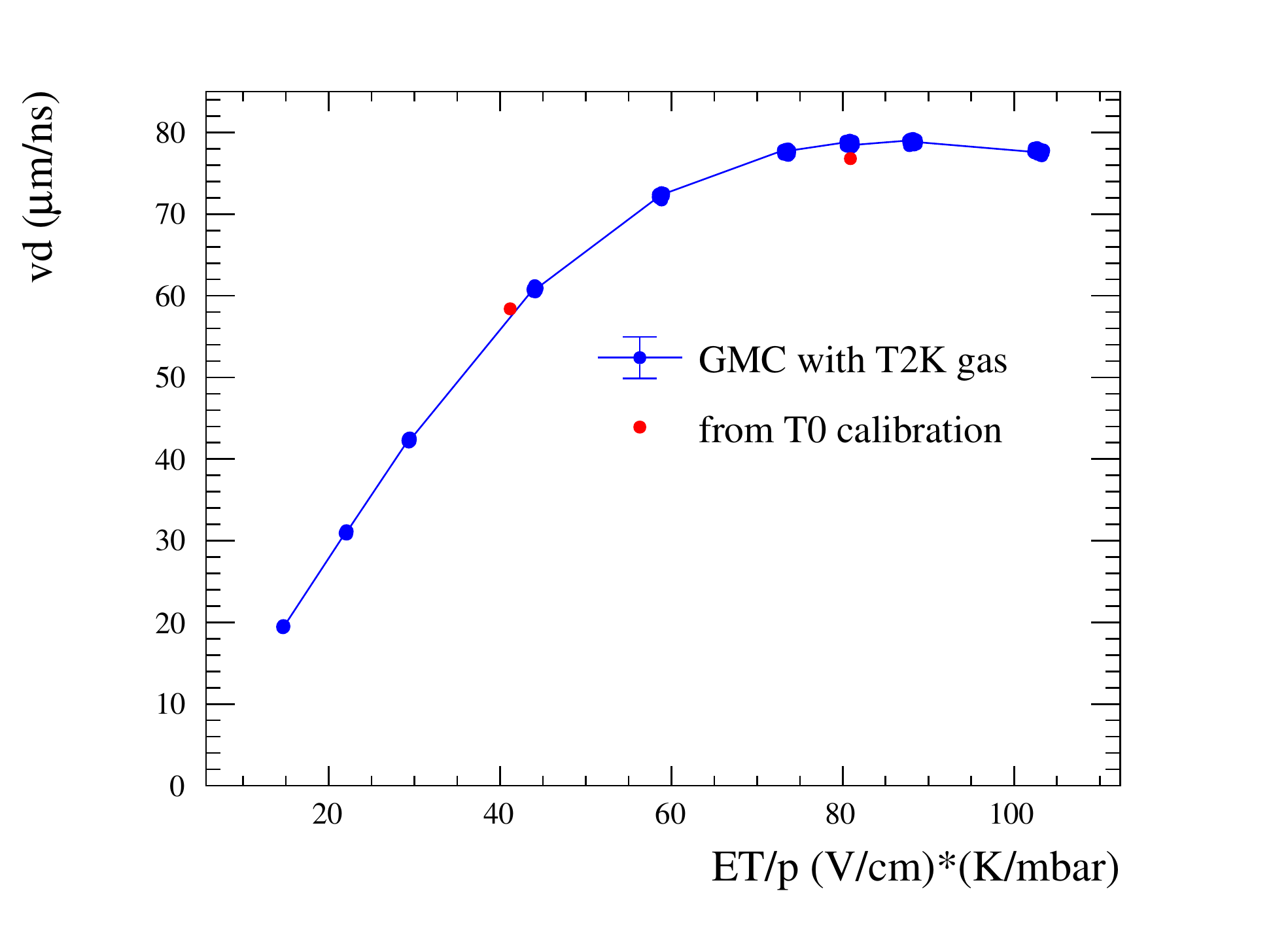}\\b
    \end{minipage}
    \caption{
       The drift velocity under a constant field of 275 V/cm as a function of days (a) and as a function of electric field (b) with density corrections applied.
       %Previously:  Drift velocity at constant field of 275 V/cm and under field variations with density corrections applied.
        The vertical lines on (a) correspond to the moments when the bottles of premixed gas were replaced. 
        Transition regions at ramp-up and between bottle changes can clearly be seen. 
        The measurements at different fields in (b) correspond to a full day during the first bottle (14.06).
    }
    \label{fig:GMC_vd}
\end{figure}

We have calibrated the $T_{0}$, the moving table position, and the drift velocities at different drift fields with a 4 GeV/c electron beam and a short peaking time of 116 ns.
At the standard T2K field of 275 V/cm, we obtained a drift velocity of $7.68\pm 0.03~{\rm cm}/\text{\textmu} \textrm{s}$ by drift distance scans with the accelerator beam.
A Gas Monitoring Chamber (GMC), identical to the ones deployed at T2K's ND280 detector \cite{Abgrall:2010hi}, monitored the exhaust gas for the duration of the beam-time.
The GMC measured a drift velocity of $7.81\pm 0.02~{\rm cm}/text{\textmu} \textrm{s}$ at the T2K field,
in agreement with the beam scans, see \autoref{fig:GMC_vd}.
Under the electric field of 140 V/cm, which is associated with the minimum transverse diffusion, we found a drift velocity of $5.84~{\rm cm}/\mu \textrm{s}$.
%previously: At the field of minimum transverse diffusion at 140 V/cm, we found a drift velocity of $5.84 {\rm cm}/\mu s$.
Despite the impact of the gas bottle changes on the gas properties, the drift velocity $V_{\rm drift}$ under the electric field of $E=275 {\rm V/cm}$ varies less than $ 6 \text{\textperthousand}$.
%Previously: Even though a difference between the premixed gases is visible, the overall stability of $V_{\rm drift}$ at the drift velocity maximum $(E=275 {\rm V/cm})$ is $\leq 6 \text{\textperthousand}$.

 % \autoref{fig:DEASY_T0} shows the drift velocity scan with different electric field. The extracted $T0$ is $4.55 \mu s$. \\

%% file: spread.tex
%editor: Samira
%%%%%%%%%%%%%%%%%%%%%%%%%
%\paragraph{Characterization of charge spreading}
%%%%%%%%%%%%%%%%%%%%%%%%%
%characterisation 

\begin{figure}[htbp]
    \centering
    \begin{minipage}{0.30\linewidth}
        \centering
        \includegraphics[width=\textwidth,origin=c,angle=0]{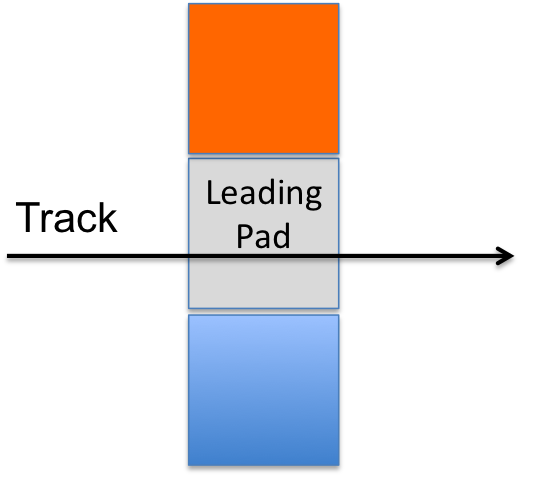}
    \end{minipage}
    \begin{minipage}{0.48\linewidth}
        \centering
        \includegraphics[width=\textwidth,origin=c,angle=0]{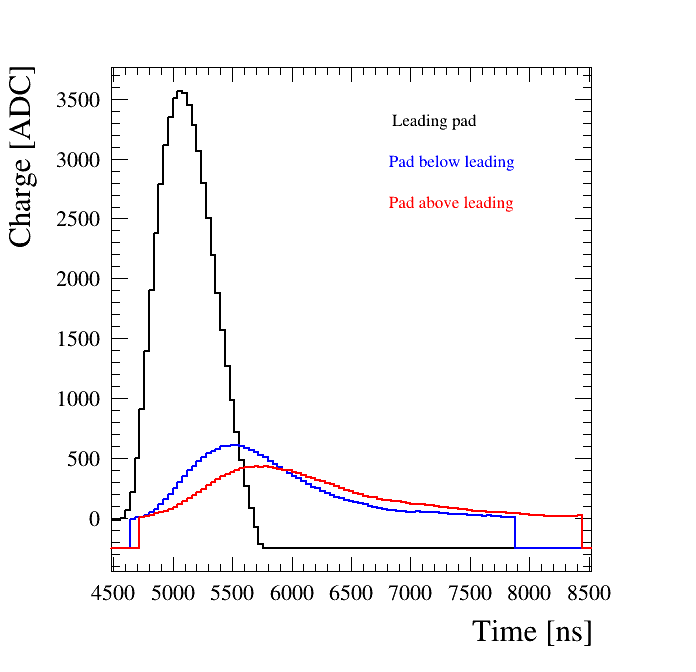}
    \end{minipage}
    \caption{A schematic view of a column cluster (left) and the waveforms of each pad composing this cluster (right) for 412 ns shaping time.
    %before: The schematic view of the ``cluster'' (left) and the signal waveforms of a given pads in a cluster (right) for 412 ns shaping time.
    }
    \label{fig:DESY_waveform}
\end{figure}

As explained before, the resistive Micromegas technology produces a spreading of the collected charge into neighbouring pads. The charge spreading phenomenon, which drives the waveform shape on the pads adjacent the "leading" ones receiving the inital charge from the avalanche, is described in~\cite{Dixit_DLC_nim2004}. The signal induced by the resistive layer is smaller in amplitude, delayed and longer in time compared to the signal from direct charge deposition from the track in the leading pad. The time delay of the waveform increases with the distance of the adjacent pad to the track. Hence, the charge spreading is significant in the transverse direction w.r.t. the track projection reconstructed in the ERAM plane, while in the longitudinal direction it is masked by the direct charge. To study the phenomenon of interest, we define a ``cluster'' as a group of pads in the perpendicular direction to the track. A schematic view of such a cluster for horizontal tracks and waveforms in the leading and in adjacent pads are shown in~\autoref{fig:DESY_waveform}.  
% Inserted the clarification why we are looking at the pads in the transverse direction w.r.t. the track.

\begin{figure}[htbp]
\centering
\includegraphics[width=.48\textwidth,origin=c,angle=0]{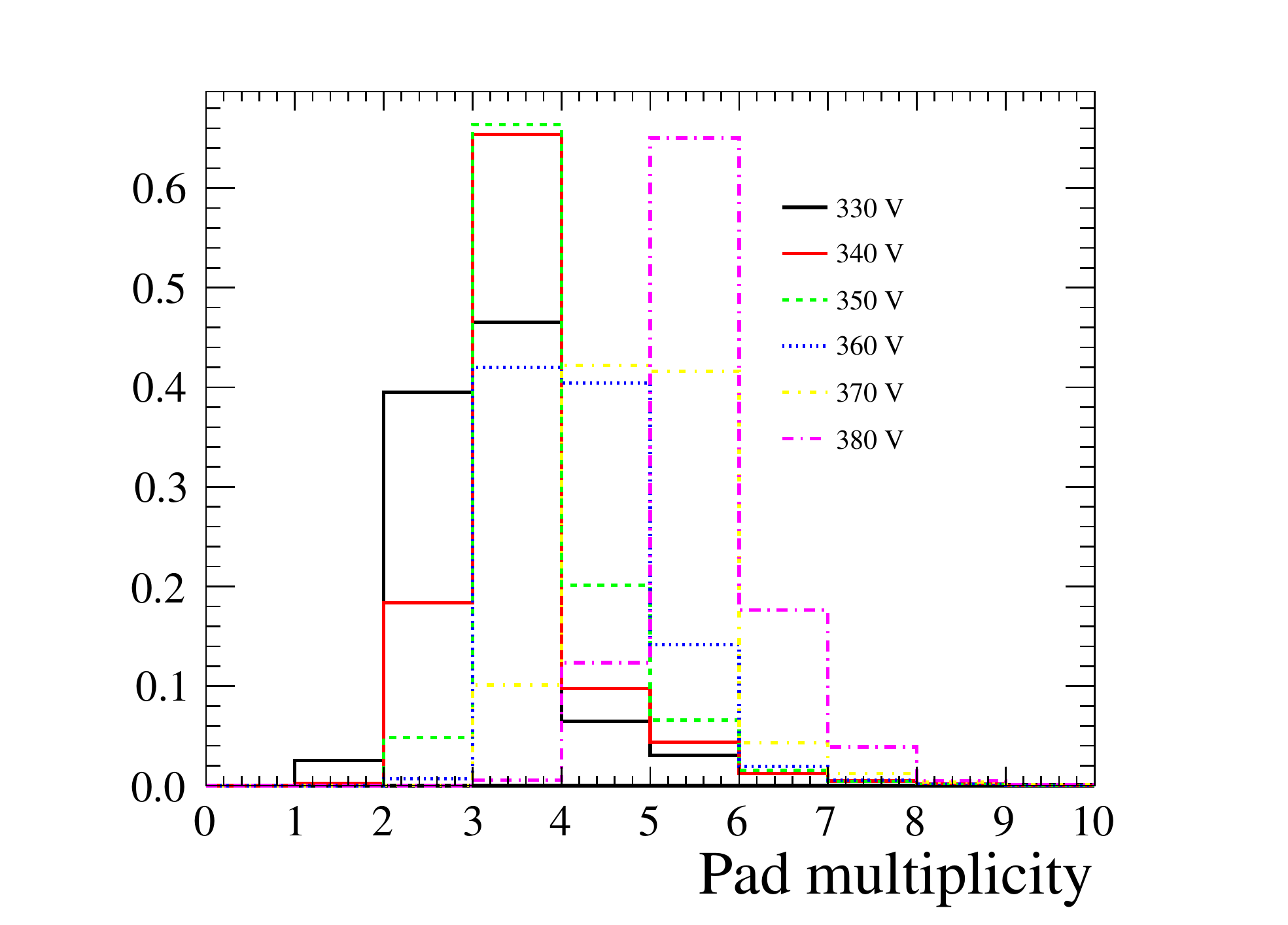}
\includegraphics[width=.48\textwidth,origin=c,angle=0]{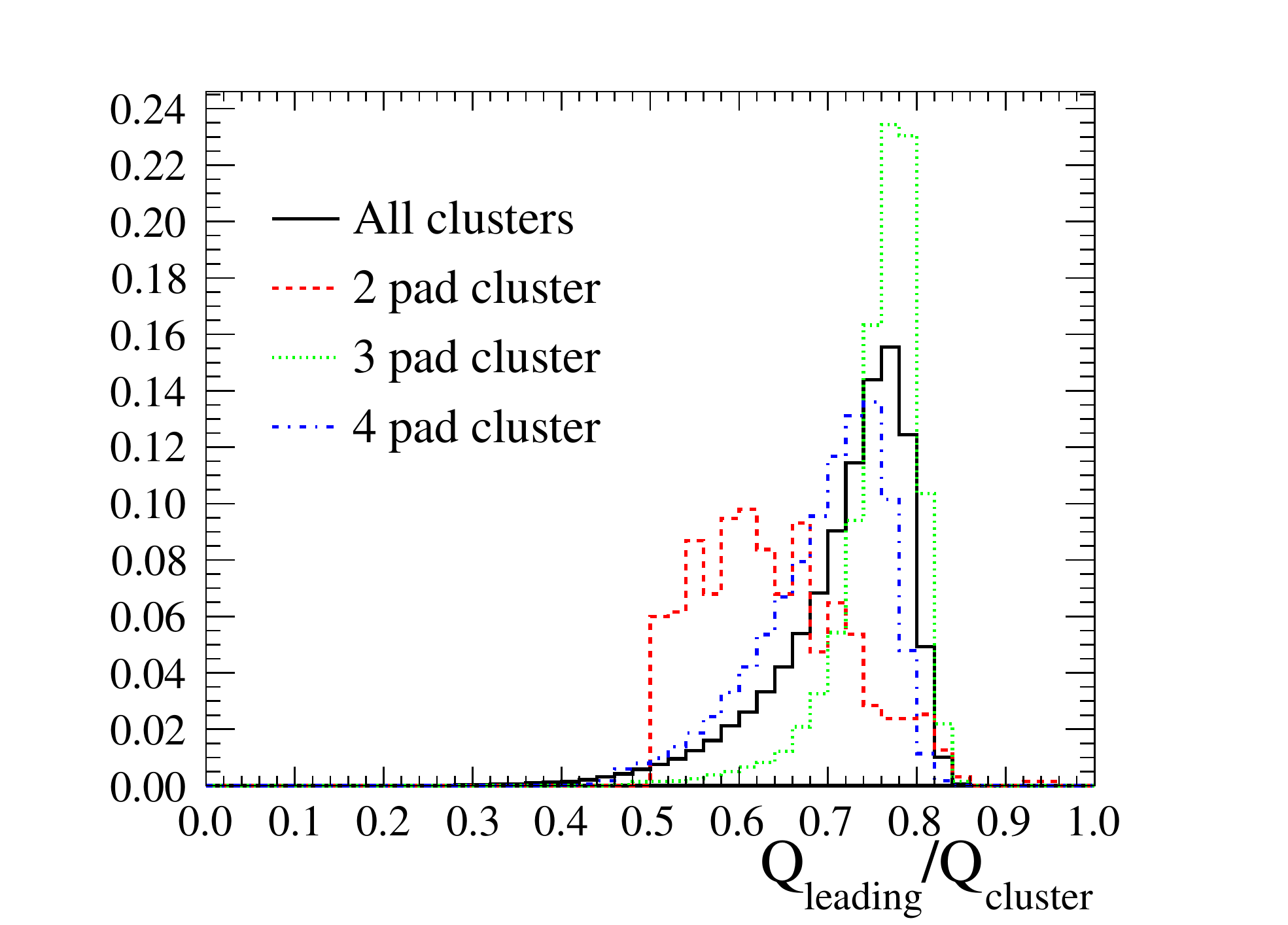}
\caption{ The number of pads in a cluster versus the DLC high voltage (left) and the fraction of the cluster charge collected in the pad with largest signal at 360 V (right).}
\label{fig:DESY_Multiplicity}
\end{figure}

% Before: The charge spreading can be clearly observed in the signal waveforms of adjacent pads, as shown in Figure~\ref{fig:DESY_waveform}.
For each pad, the maximum of the waveform is used as an estimator of the charge, which corresponds to different times for the leading and adjacent pads. In \autoref{fig:DESY_Multiplicity}, the pad multiplicity per cluster and the ratio of the charge measured in the pad with the largest signal with respect to the sum of the charges of the pads in the cluster  ($q_\textup{max}/q_\textup{cluster}$) are shown. 

Most of the clusters are formed by more than two pads and the pad with the largest signal contains typically 80\% of the total charge of the cluster. 
The effect of the high voltage on the multiplicity is also clearly seen in \autoref{fig:DESY_Multiplicity}. 
The cluster multiplicity increases with the high voltage, as the probability for the smaller signals in some pads to  pass the threshold increases with the gain.
%%%%%%%%%%%%%%%%%%%%%%%%%%%%%%
%%%%%%%%%%%%%%%%%%%%%%%%%
%%%%%%%%%%%%%%%%%%%%%%%%%
%%%%%%%%%%%%%%%%%

\begin{figure}[!ht]
    \centering
    \begin{minipage}{0.24\linewidth}
        \centering
        \includegraphics[width=\linewidth]{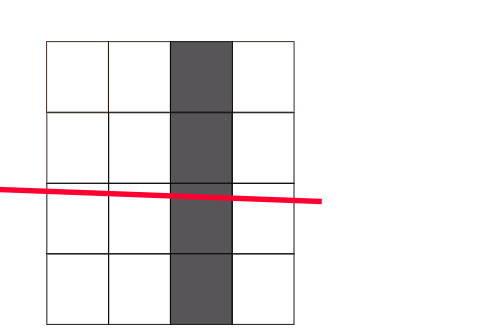} \\ a
    \end{minipage}
    \begin{minipage}{0.24\linewidth}
        \centering
        \includegraphics[width=\linewidth]{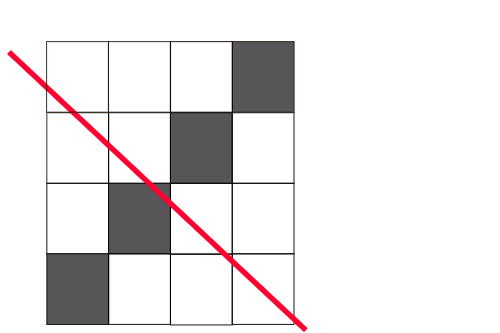} \\ b
    \end{minipage}
    \begin{minipage}{0.24\linewidth}
        \centering
        \includegraphics[width=\linewidth]{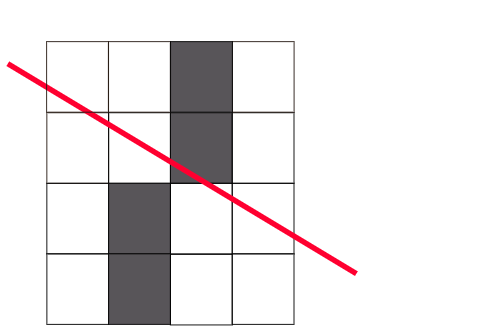} \\ c
    \end{minipage}
    \begin{minipage}{0.24\linewidth}
        \centering
        \includegraphics[width=\linewidth]{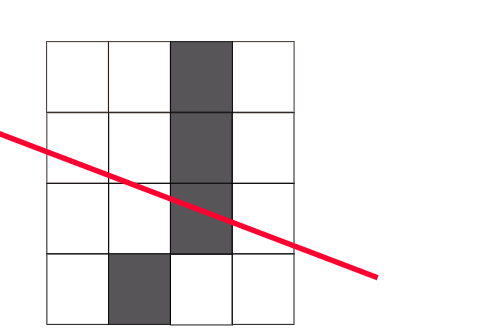} \\ d
    \end{minipage}
    \caption{The different cluster patterns that can be used depending on the track (red line) angle with the pads sides : (a) column, (b) diagonal, (c) 2 by 1, (d) 3 by 1. For each case, the colored pads correspond to one cluster, associated to the colored leading pad crossed by the track. 
    %before: The cluster of pads used to extract the transverse charge spreading for the tracks of different angles (shown in red line): (a) column, (b) diagonal, (c) 2 by 1, (d) 3 by 1.
    }
    \label{fig:spatial:clusters}
\end{figure}

For tracks projections in the pad plane which are parallel to the sides of the pads, transverse spreading is defined precisely within the given column. While for inclined tracks and large square pads, the separation between longitudinal (along the track) and transverse spreading is more complicated. In order to distinguish these two spreading topologies, we define more sophisticated cluster patterns shown in \autoref{fig:spatial:clusters}. These clusters are repeated to pave the whole ERAM. In the case of square pads, these patterns are optimized for angles with tangents 0 (column), 1 (diagonal), 0.5 (2 by 1), 0.3 (3 by 1) respectively. In the case of rectangular pads, the optimal angles are slightly different. 

%% file: reco.tex
% !TEX root = resistiveMM.tex
%Editor: Sergey, Marion

%comment: Maybe we should use the present tense no? or the past perfect at least no?
%Previously: we focused
In the test beam analysis, we focus on the studies of the through-going tracks as more complicated typologies (e.g. showering, curved low-energy tracks) are difficult to interpret. Hence simple reconstruction algorithms like  DBSCAN~\cite{Ester96adensity-based} are sufficient in our case. 
 
\begin{figure}[!ht]
    \centering
    \begin{minipage}{0.49\linewidth}
        \centering
        \includegraphics[width=\linewidth]{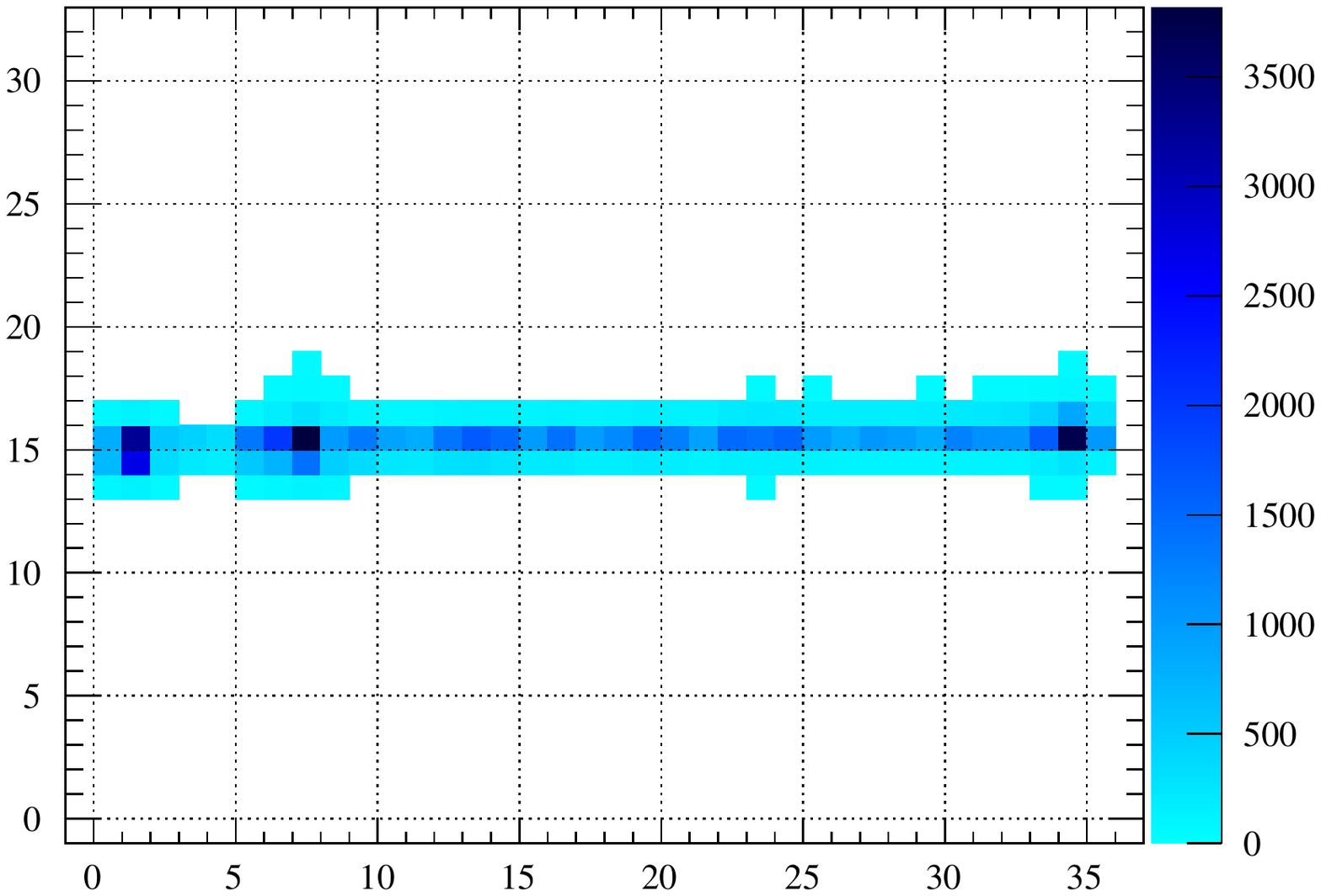} \\ a
    \end{minipage}
    \begin{minipage}{0.49\linewidth}
        \centering
        \includegraphics[width=\linewidth]{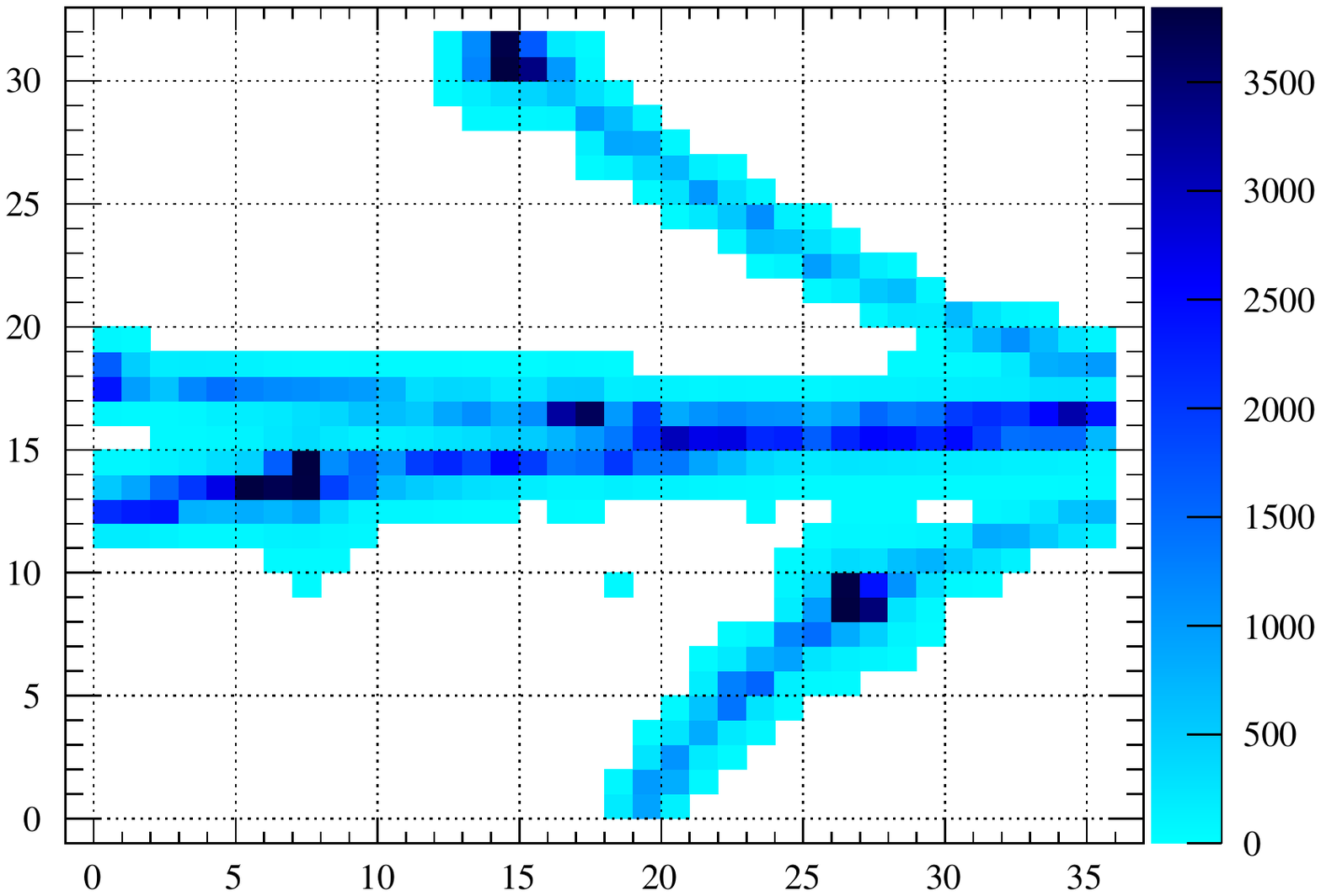} \\ b
    \end{minipage}
    \caption{Event displays of the (a) single track and (b) multi-track in the prototype.}
    \label{fig:reco:event_topo}
\end{figure}

We select a track if it is crossing the whole detector without breaks or splits. A split is defined as the case where there is more than one cluster in a given column. An event containing a split is thus a multiple track candidate and is rejected in our analysis. 
%Comment:can maybe be even clearer
%Previously: Splits are recognized as more than 1 cluster in the given column that corresponds to a multi-track event. 
However, with the resistive spreading, two close parallel tracks may not be separated by a gap and thus misreconstructed as one single track. 
To reject such a topology, a cut on the pad multiplicity in each cluster was implemented. 
The cut value was optimised for each Micromegas voltage and electronics shaping time.
%Previously: However, with the resistive spreading, two collinear tracks may not be separated with a gap, thus misreconstructed as a single one. To reject such a topology, the cut on the pad multiplicity in the cluster was implemented. The cut value was optimised for the MicroMegas voltage and electronics shaping time.
Examples of the accepted and rejected events are shown in \autoref{fig:reco:event_topo} (a) and (b) respectively. 
%Previously: The event display examples of the track of interest and dismissed event are presented in~\autoref{fig:reco:event_topo} (a) and (b) respectively. 
%Comment: can maybe be even clearer, a little heavy with of..of...
%Proposal
%We have to add a multiplicity cut to remove residual problematic multiple track events which could not be detected with a split. For instance, events with a very vertical secondary track. 

%% file: spatial.tex
% !TEX root = resistiveMM.tex

%Editor: Sergey, Marion
\renewcommand{\degree}{${}^\circ$}

% Claudio proposal:
As described in~\autoref{sec:introduction}, the momentum resolution has to be better than 10\% at 1 GeV/c but, to fully exploit the ND280 upgrade capabilities, an even better momentum resolution would be desirable. The momentum resolution is directly connected to the spatial resolution obtained for one cluster through the Gluckstern formula~\cite{GLUCKSTERN1963381}. For tracks with 70 point measurements (using clusters), a maximum drift distance of 1 m, and a magnetic field of 0.2~T, a spatial resolution of $\sim800 \text{\textmu} \textrm{m}$ would be sufficient to reach a momentum resolution of 10\% at 1 GeV/c. 

As we will show in this section, the resistive Micromegas technology allows one to significantly improve the spatial resolution with respect to the bulk Micromegas, even in presence of slightly larger pads thus allowing one to reduce by $\sim30\%$ the total number of electronic channels for the same active surface. 
The test beam data have been used to characterize the ERAM module performances for electrons with different angles with the pads sides.

\subsection{The Pad Response Function method}

The charge spread described in the~\autoref{sec:spread} results in charge detection in a few pads around the avalanche arrival point. The charge measurements are discrete with the finite pad size, while the spreading in the RC layer is continuous. Moreover, the signal induced in the adjacent pads does not result from a real sharing of an initial charge, and it is measured at different times, even if it is related to the charge measured in the leading pad. Thus the barycentric method (Center of Charge, weighted mean) that assigns all the collected charge to the pad centre doesn't provide a precise position reconstruction. Instead, we used the so-called Pad Response Function (PRF) which characterizes the relation between observed charge ratios and track position w.r.t. the pad (\autoref{eq:spatial:PRF}). 
This method improves the spatial resolution compared to the barycentric method for TPCs with the resistive anode~\cite{Attie:2011zz}. The PRF is defined as:

\begin{equation}
    PRF(x_{track}-x_{pad}) = Q_{pad}/Q_{cluster}
    \label{eq:spatial:PRF}
\end{equation}
where $x_{track}$ is the reconstructed position of the track, $x_{pad}$ is the centre of the pad, $Q_{pad}$ is the charge collected on a given pad and $Q_{cluster}$ is the charge collected on the whole cluster. The definition of the cluster is the same as described in~\autoref{sec:spread}: it's a group of pads where one receives a charge from the initial avalanche and the others detect the charge spread in the resistive foil.

To parametrize the PRF we used the empirical ratio of two symmetric \text{$4^{th}$} order polynomials proposed in~\cite{Boudjemline:2006hf}:

\begin{equation}
PRF(x, \Gamma, \Delta, a, b) = A \times \dfrac{1+a_{2}x^{2}+a_{4}x^{4}}{1+b_{2}x^{2}+b_{4}x^{4}}
\label{eq:spatial:prf_ana}
\end{equation}
where the parameters $a_{i}$ and $b_{i}$ can be related to the more physical parameters: the full width at half maximum $\Gamma$, the base width $\Delta$, and two scaling parameters $a$ and $b$.

\subsection{Spatial resolution estimation}

\begin{figure}[!ht]
    \centering
    \begin{minipage}{0.49\linewidth}
        \centering
        \includegraphics[width=\linewidth]{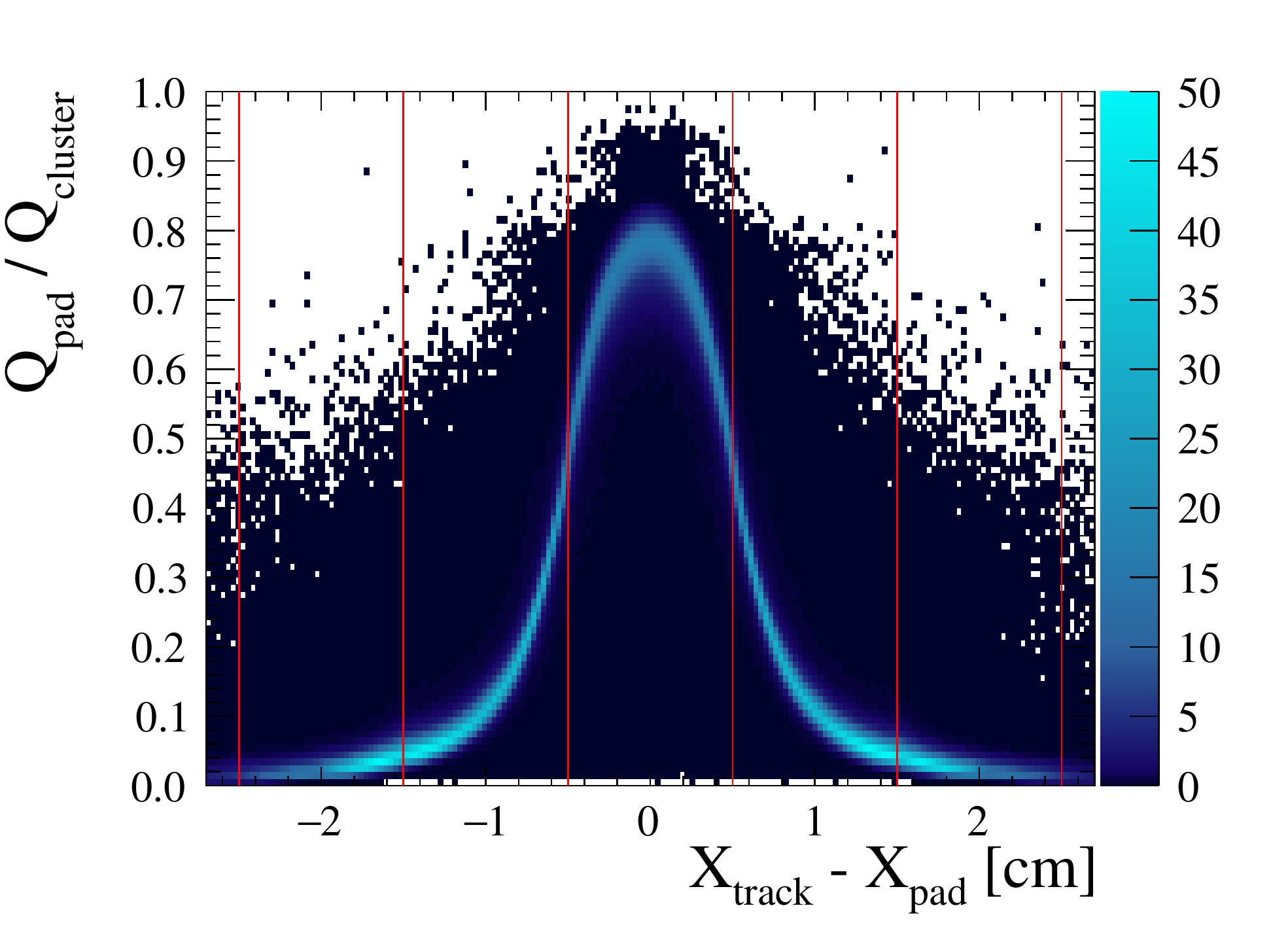} \\ a
    \end{minipage}
    \begin{minipage}{0.49\linewidth}
        \centering
        \includegraphics[width=\linewidth]{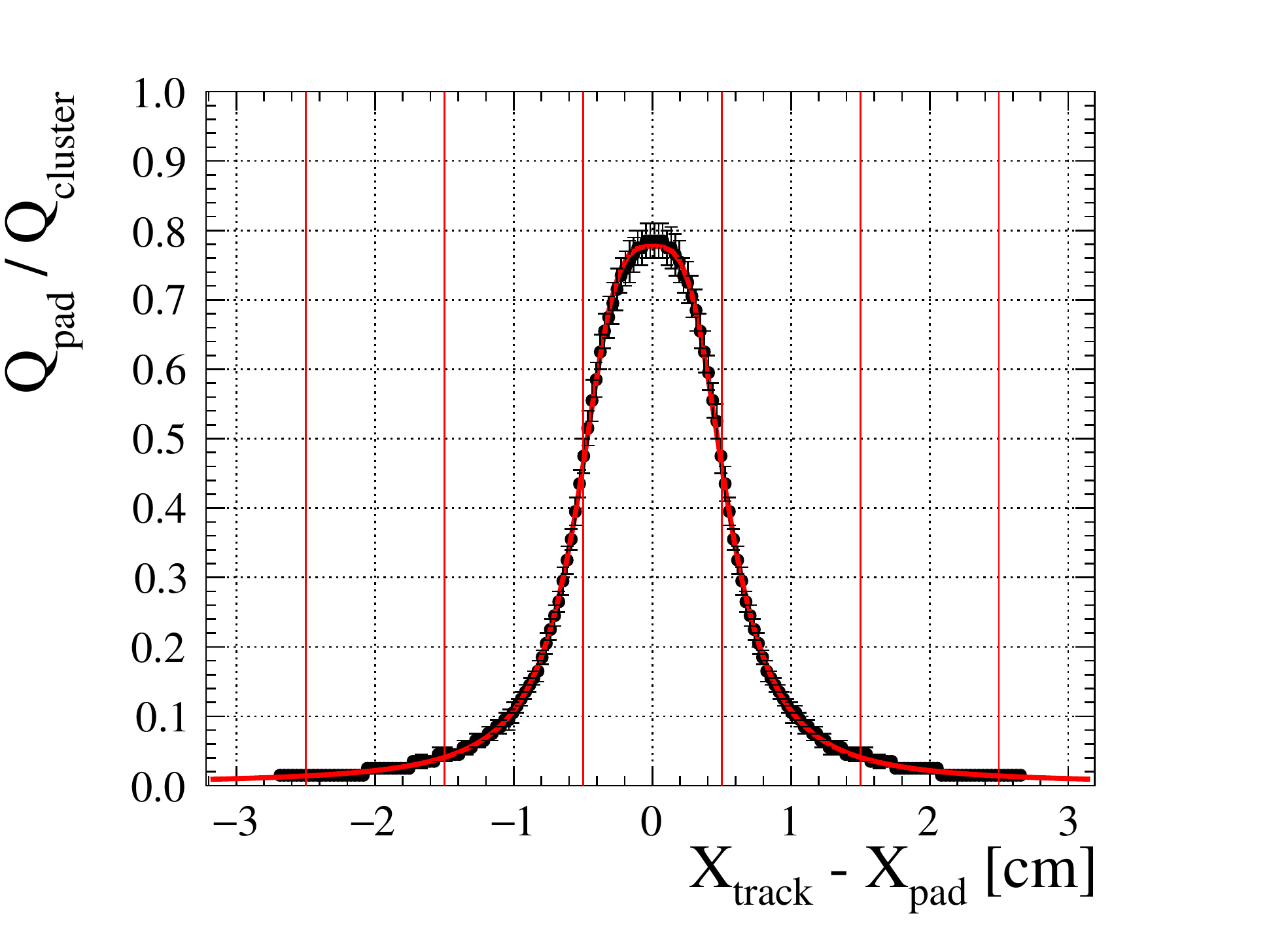} \\ b
    \end{minipage}
    \caption{The Pad Response Function (PRF) obtained with (a) scattered plot and (b) results of the profile and fit with analytical function. The pad borders are represented by vertical lines.}
    \label{fig:spatial:prf}
\end{figure}

This parametrized function of the PRF allows one to determine the parameters using an iterative method.
To get a first guess of the PRF parameters, we use the track position reconstruction obtained from the barycentric method. After having defined the position of all the clusters, the global track is fit with a parabola. The fit is based on many measurements using the different clusters along the track ($\geqslant34$) thus it is considered as a true track position and the PRF scattered plot is filled (\autoref{fig:spatial:prf} (a)). 
%comment: maybe do 2 sentences
The scattered plot is profiled along Y axis to form a graph that is further fit with the analytical function from \autoref{eq:spatial:prf_ana} (\autoref{fig:spatial:prf} (b)).

After having determined the PRF, the track position in each cluster is obtained with the following $\chi^2$ minimization:

\begin{equation}
    \label{eq::spatial:chi2}
    \chi^2=\sum_{pads}\frac{Q_{pad}/Q_{cluster} - PRF\left(x_{track} - x_{pad}\right)}{\sigma_{Q_{pad}/Q_{cluster}}}
\end{equation}
where $\sigma_Q$ is the uncertainty on the charge measurements. In our analysis, we assume that charge measurement probability follows a Poisson distribution, hence: $\sigma_{Q_{pad}/Q_{cluster}}=\sqrt{Q_{pad}/Q_{cluster}}$. 
We proceed through the iterative process of the PRF estimation until the track fit quality is not improving anymore. Typically this procedure converges after few iterations and for the results shown in this paper 10 iterations were used.

The spatial resolution is defined as the difference between the reconstructed position in a given cluster and the track global fit (residual). The particular cluster where the resolution is studied is excluded from the fit to prevent biases. The residuals distribution is fit with a Gaussian function whose standard deviation defines the spatial resolution.

\subsection{Spatial resolution dependence on the drift distance, momentum, high voltage}
\begin{figure}[!ht]
    \centering
    \begin{minipage}{0.49\linewidth}
        \centering
        \includegraphics[width=\linewidth]{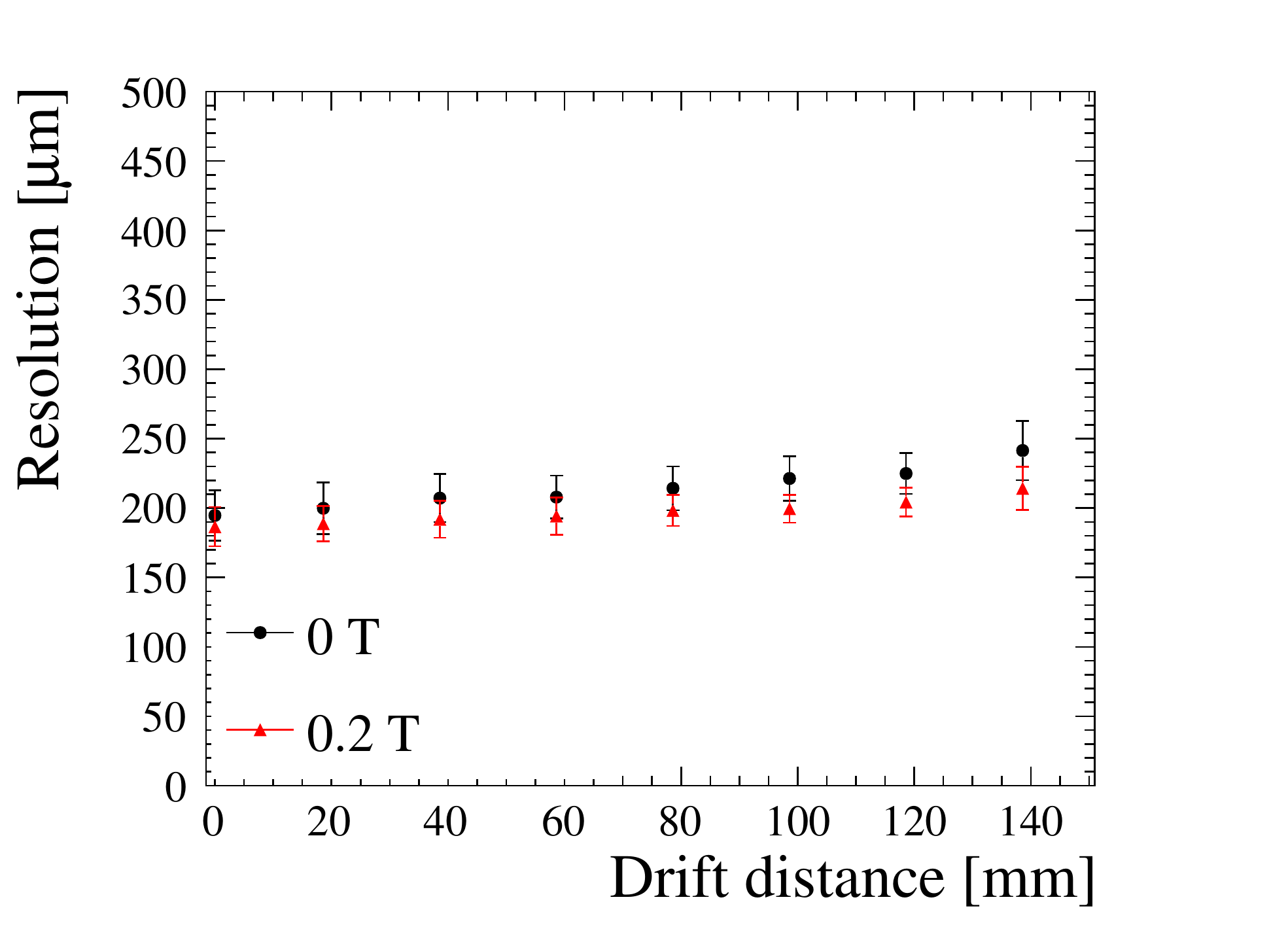} \\ a
    \end{minipage}
    \begin{minipage}{0.49\linewidth}
        \centering
        \includegraphics[width=\linewidth]{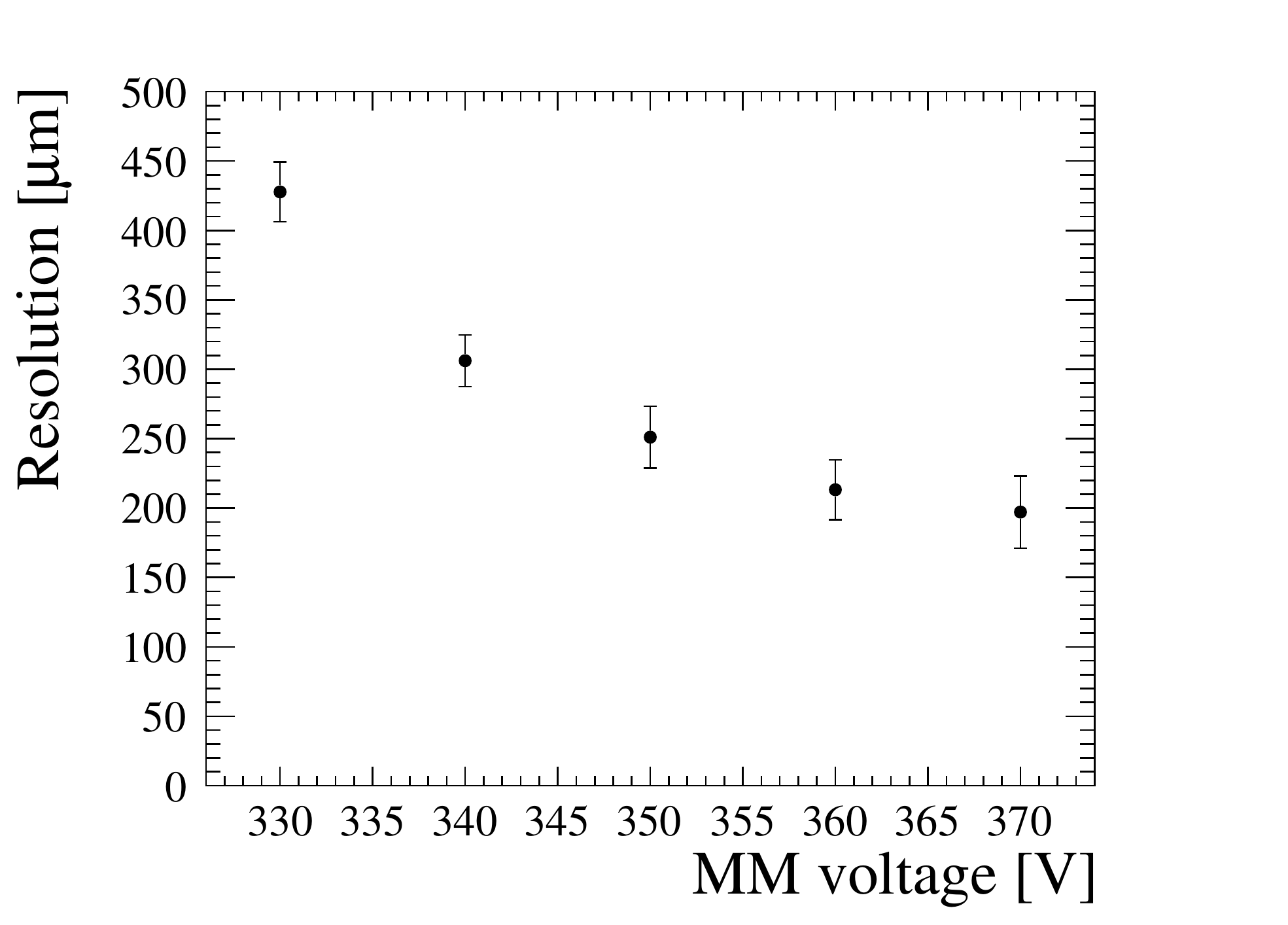} \\ b
    \end{minipage}
    \caption{Spatial resolution w.r.t. (a) beam injection position and (b) Micromegas voltage for horizontal tracks parallel to the MM plane. Points represent the mean value over detector columns and errors represent the fluctuations (RMS).}
    \label{fig:spatial:drift}
\end{figure}

The spatial resolution was studied over different samples. The beam position was varied within the drift distance of the field cage keeping the tracks parallel to the Micromegas plane. For resistive Micromegas, the spatial resolution is expected to degrade slightly for a larger drift distance affected by the transverse and longitudinal diffusion. The observed dependence, for a DLC voltage of 360 V, is shown in \autoref{fig:spatial:drift} (a). A resolution between 200 and 250 $\text{\textmu} \textrm{m}$ is observed for the whole drift length. 

In addition, the high voltage applied to the Micromegas mesh was varied to study the detector performance in different regimes. Higher voltage is expected to enhance the initial avalanche, thus increase the smaller charge spreading signal as well. Signals in the neighbour pads are then more likely to pass above the threshold and are less affected by statistical fluctuations. As track position reconstruction relies on the charge spreading measurements we expect better performance with higher voltage. 
The results are shown in~\autoref{fig:spatial:drift} (b).

\begin{figure}[!ht]
    \centering
    \includegraphics[width=0.5\linewidth]{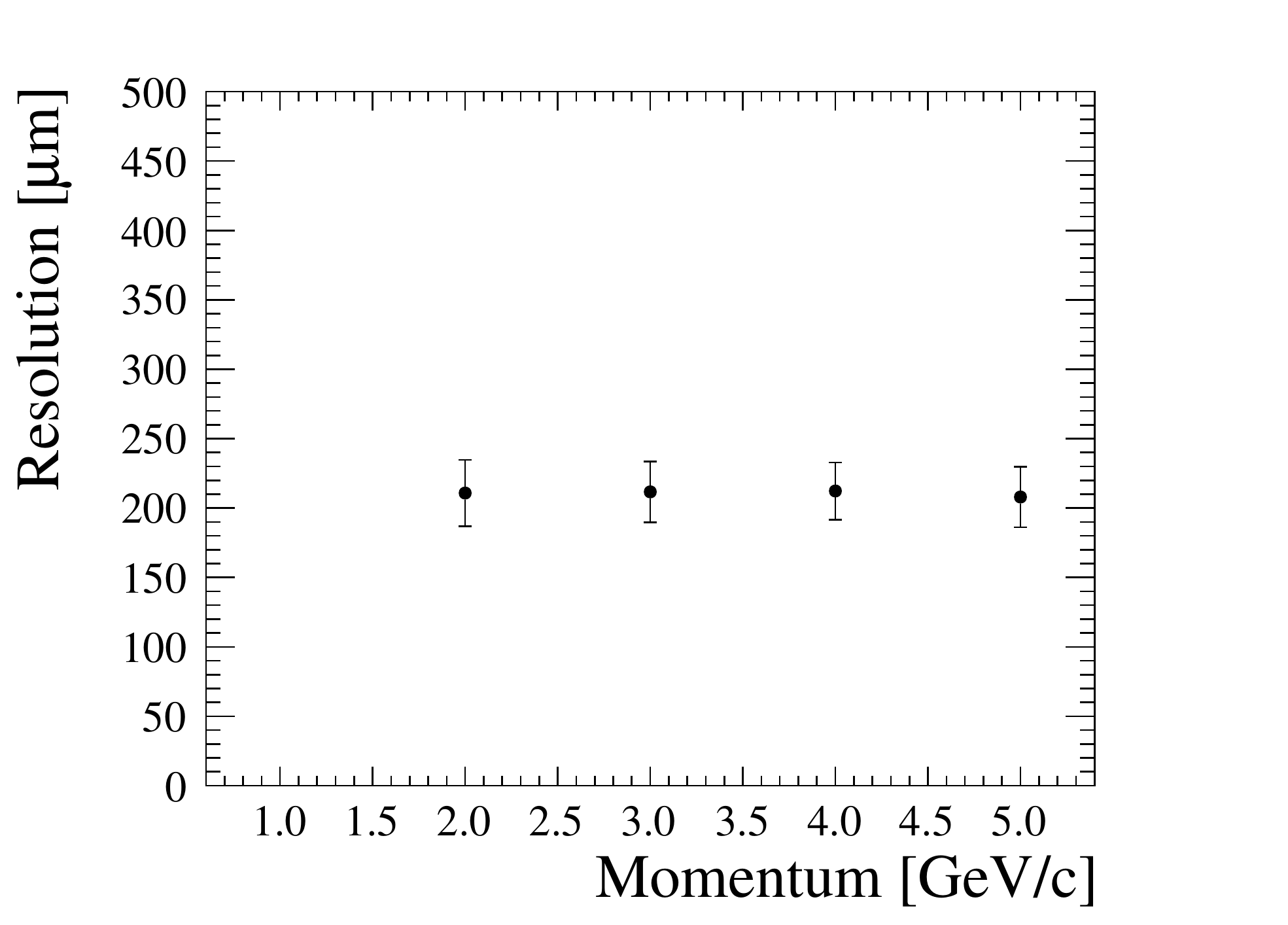}
    \caption{Spatial resolution as a function of the momentum of the electron beam.}
    \label{fig:spatial:mom}
\end{figure}

The DESY beamline allows changing the momentum of the electrons delivered to the test beam area. In this way, we studied spatial resolution as a function of the track kinematics. We find no significant changes for the position accuracy reconstruction for the tracks in the range between 1 and 5 GeV/c (\autoref{fig:spatial:mom}).

\subsection{Spatial resolution dependence on the track inclination}
Inclined tracks are expected to be reconstructed less precisely compared to horizontal ones. As an example, in the current ND280 TPCs, the resolution degrades as a function of the track angle from $600~\text{\textmu} \textrm{m}$ to $\sim1~\textrm{mm}$~\cite{Abgrall:2010hi}. It is then particularly interesting to investigate the behaviour of the spatial resolution in the ERAM detector as a function of the angle of the reconstructed track on the ERAM plane with respect to the ERAM side (track inclination).

\begin{figure}[!ht]
    \centering
    \includegraphics[width=0.8\linewidth]{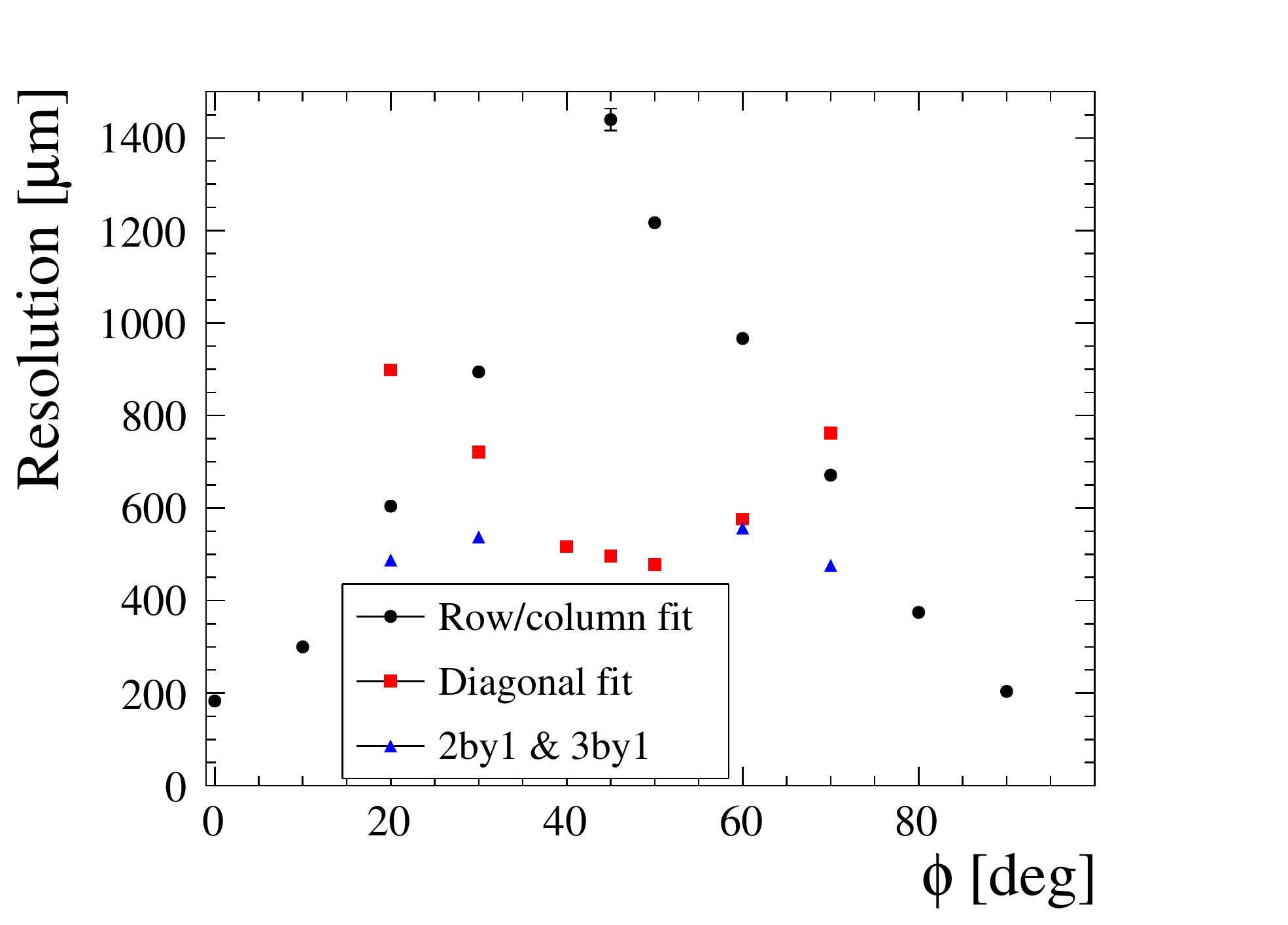}
    \caption{Spatial resolution w.r.t. track angle within the MM side obtained with different cluster definitions.}
    \label{fig:spatial:sloped}
\end{figure}

In order to do this, the different cluster patterns described in~\autoref{sec:spread} were used for different track inclination. The $\chi^2$ fit (\autoref{eq::spatial:chi2}) is applied to each cluster to extract the track position. As for horizontal tracks, the positions in given clusters are fit together to form a global fit and the iterative analysis is applied: the barycentric estimation is used as a prior following with PRF calibration. The results of the inclined track spatial resolution estimations are shown in \autoref{fig:spatial:sloped}.

% results

As expected, the column clustering method leads to a severe spatial resolution degradation with the track slope, reaching a maximum of $1.4~mm$ for 45\degree~tracks. 
For these angles, the diagonal pattern provides a significant performance improvement. In the intermediate regions (20\degree, 30\degree, 60\degree, 70\degree), the best result is achieved with the more complex patterns: ``2 by 1'' and ``3 by 1''. The asymmetry w.r.t. 45\degree~is caused by the rectangular pad shape 11.3$\times$10.2 mm. Thus, the diagonal pattern is considered a better choice for 48\degree~tracks than for 45\degree~tracks. 
Hence, tracks inclined with 50\degree~are reconstructed more accurately compared to 40\degree~tracks. Similar behavior is observed for all the other patterns.

By taking the best clustering algorithm we observe a spatial resolution better than $600~\text{\textmu} \textrm{m}$ for all the angles. We understand the difference between horizontal tracks and inclined tracks as due to the larger effective pad size for diagonal clustering and to the rectangular shape of the pads, while diagonal clustering would work better for square pads.
For a spatial resolution of $600~\text{\textmu} \textrm{m}$, for 70 point measurements and a magnetic field of 0.2~T we expect a momentum resolution of 6\% at 1 GeV/c that scales linearly with the spatial resolution.

\subsection{Bias measurements}

\begin{figure}[!ht]
    \centering
    \includegraphics[width=\linewidth]{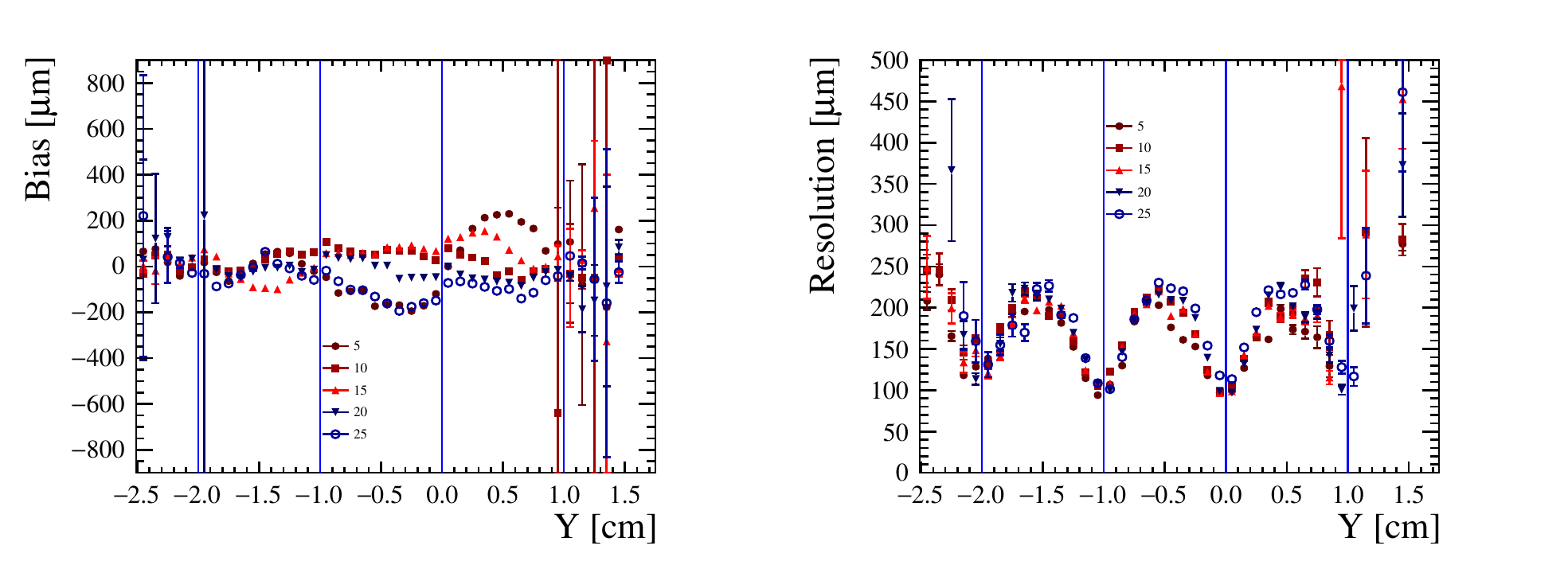}
    \caption{The resolution and the bias of the track reconstruction over the position in the pad for various columns. The pad borders are represented with vertical lines. In this coordinate system, the beam is centered around $Y=-0.5cm$, but tracks are also measured on the neighbour pads due to the beam spread. }
    \label{fig:spatial:bias}
\end{figure}

As described above, we define the spatial resolution as a standard deviation of the difference between the reconstructed position in a given cluster and a global track fit. Meanwhile, the mean value of the residuals is also an important characteristic that shows the bias of our measurements. In particular, it is interesting to study the biases with respect to the track position in the pad. For that, we use the natural beam spread. The electron beam profile is nearly Gaussian with a standard deviation $\approx$ 1 cm. We sample the residuals with the reconstructed track position in the pad. Thus, we can study the resolution and biases in the different pad regions. In \autoref{fig:spatial:bias}, we show both the spatial resolution and bias per column. Individual PRFs were used for each column to analyze the behaviour in the different regions of the detector independently. The resolution undergoes some oscillations because if a track is close to one of the leading pad borders, the neighbouring pads see a larger signal and we thus have a more reliable input for the position reconstruction.

\begin{figure}[!ht]
    \centering
    \includegraphics[width=0.6\linewidth]{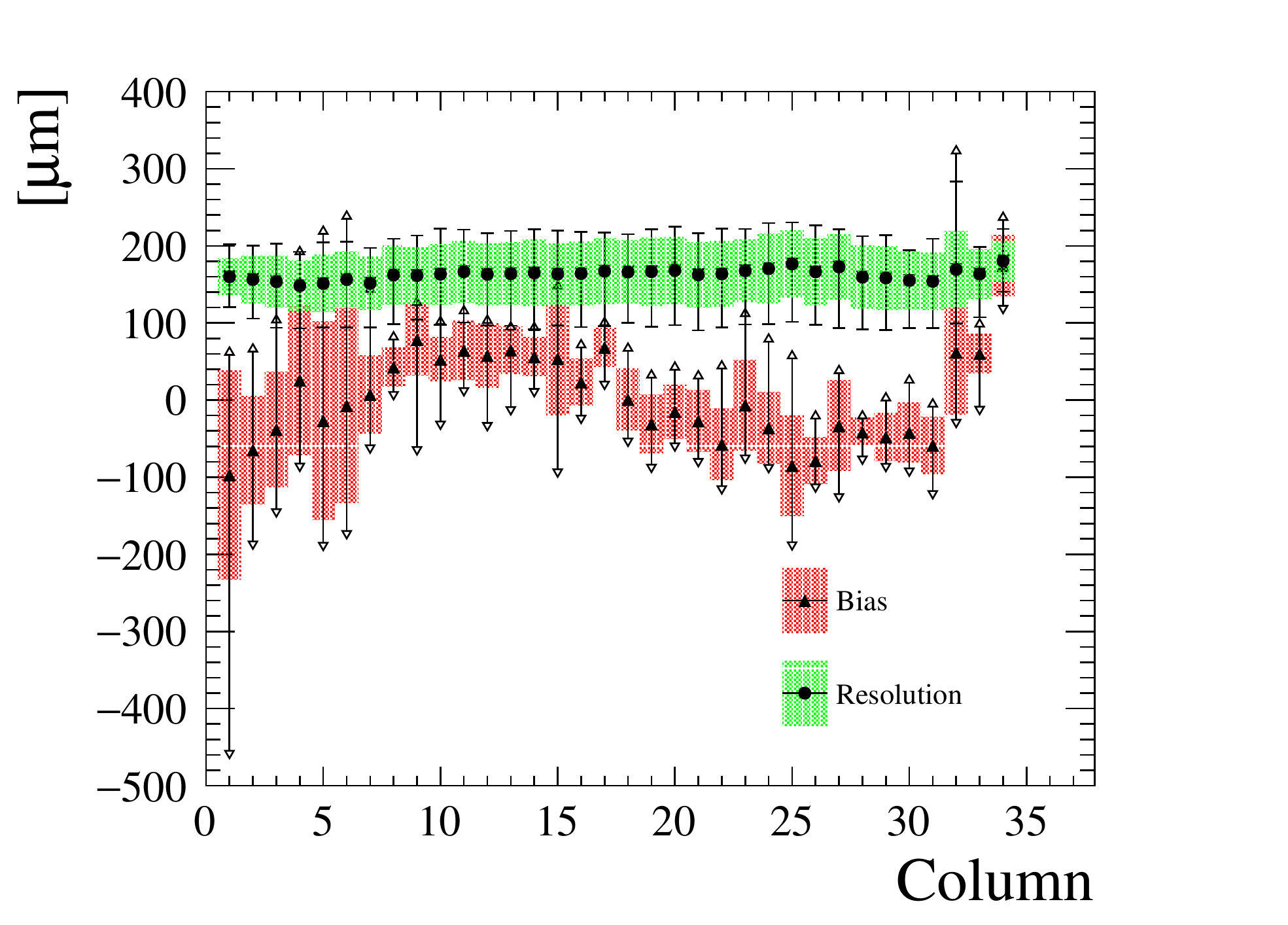}
    \caption{The spatial resolution and bias fluctuations observed for the different position of the track in the pad. Dots represent the mean value in a given column, filled areas correspond to the RMS, and error bars represent minimum and maximum values.}
    \label{fig:spatial:bias_all}
\end{figure}

We generalize the bias study for the whole detector. \autoref{fig:spatial:bias_all} represents the fluctuations of the resolution and biases in the given column. 
We conclude that for most of the detector the biases are under control and smaller compared to the spatial resolution. In the downstream part of the detector, we found larger biases that could be related to the non-uniformities in the resistivity of this ERAM detector that will be described in \autoref{sec:rcmap}.
%other idea: the inhomogeneity of DLC layer resistivity

%% file: dedx2.tex
One of the main goals of a TPC is to perform particle identification (PID) based on the measurement of the ionization produced by charged particles crossing the gas volume. 
%Previously: One of the main role of a TPC is to perform particle identification based on the measurement of the ionization produced by charged particles crossing the gas volume. 
The PID capabilities depend on the resolution in the ionization energy loss measurements. 
%Previously: Such resolution depends on the number of independent measurements of the ionization (i.e. the number of clusters) and on the amount of ionization in each cluster. 

In the case of T2K TPCs, the PID is mainly used to distinguish electrons (produced by $\nu_e$) from muons (produced by $\nu_{\mu}$). 
In the momentum range studied by T2K, the amount of ionization between electrons and muons differs by $\sim40\%$.
Therefore, a resolution of less than 10\% is needed to efficiently distinguish these two particles.
%Previously: In the region of interest for T2K, the amount of ionization between electrons and muons differs by $\sim40\%$ and a resolution better than 10\% is required. 
In general, the resolution depends on the number of independent ionization measurements (i.e. the number of clusters) and on the amount of ionization in each cluster. 
For the existing TPCs, a resolution of $8\%$ was obtained by combining measurements in 2 Micromegas detectors (72 independent measurements of ionization). 

In this section, we will describe the performances observed with one single ERAM detector (36$\times$32 pads).
%Previously: In this section, we will describe the observed performances with one single ERAM detector (36$\times$32 pads).

%Indeed to truncate the high energy ionization fluctuations,
The method used to estimate the energy loss of a given track is called the truncated mean method: the charges contained in each cluster of the track are sorted by increasing order and only a fraction of the lowest charges is kept to compute the mean deposited energy per track. 
%Previously: The method used to estimate the energy loss is the truncated mean method. In this method, the charge contained in each clusters of one track are ordered as a function of the measured charge and only a fraction of them is used to compute the mean deposited energy.
%comment: dE/dx defined for each track
Such a method allows to reject clusters with a large amount of charge, due to fluctuations in the ionization processes, that would degrade the relative resolution on the mean value, and thus the power to separate different types of particles. 
%Indeed, the high energy clusters, due to fluctuations in the ionization processes, are removed, to prevent resolution deterioration. 

\begin{figure}[H] 	
\centering
\includegraphics[scale = 0.3]{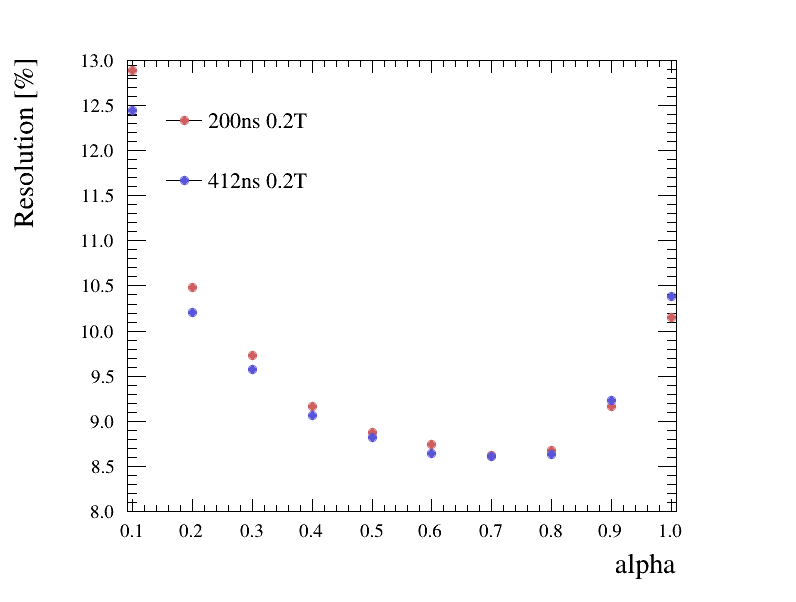}
\caption{The dependence of the $dE/dx$ resolution on the truncation factor $\alpha$ for the beam parallel to the pad sides with a magnetic field of 0.2 T and electronics peaking times of either 200 ns and 412 ns}
\label{fig:dedx_truncation_trunc}
\end{figure}

The dependence of the $dE/dx$ resolution on the truncation fraction is shown in \autoref{fig:dedx_truncation_trunc}. 
%Previously: The dependence of the resolution on the truncation factor is shown in \autoref{fig:dedx_truncation_trunc}. 
The best resolution is obtained for values of truncation fraction between 50\% and 70\%. 
Therefore, for all the results presented in this section, a truncation factor of 70\% is used.
%Previously: For all the results presented in this section, a truncation factor of 70\% is used.

\subsection{Definition of the cluster charge}
%\subsection{Definition of the charge per cluster}
%Before: \subsection{Definition of charge per cluster}

As explained before, the basic ingredient of the $dE/dx$ resolution is the amount of charge seen in each cluster. This quantity can be defined in different ways. 
%Before: As explained before, the basic ingredient of the deposited energy resolution is the amount of charge deposited in each cluster. This quantity can be defined in different ways. 
%comment: the charge is not deposited, but induced

In the existing ND280 TPCs or in the results published in~\cite{Attie:2019hua}, the cluster charge is defined by summing the waveform maximum seen in each pad composing the cluster.
%In the existing ND280 TPCs or in the results published in~\cite{Attie:2019hua}, the charge is defined by summing the maximum of the waveform in each pad of the cluster (Q$_{sum}$).
This definition was shown to be a good estimator of the charge in the case of bulk Micromegas, or when the drift distance is large enough to allow the predominance of transverse diffusion over the charge spreading induced in the ERAM module.
%Before: This is a good estimator of the charge with bulk MicroMegas (as in the case of the ND280 TPCs) or when the drift distance is large enough that the transverse diffusion dominates over the spreading induced by the ERAM module.
%comment: sounds strand to me

However, in the present configuration, the limited size of the TPC implies that the transverse diffusion is small and that charge spreading dominates, except when the track is close to the pad border. 
Therefore, by summing the maximum of the waveform seen in each pad of the cluster, we are double counting the leading pad charge: first, it is seen in the leading pad before the spread, and then seen by the neighboring pads after the spread. 
%before: In the present configuration, instead, the transverse diffusion is small due to the small size of the TPC and, by summing the maximum of the waveform in each pad of the cluster, we are double counting the same charge as seen by the leading pad, before the spread, and by the neighboring pads, after the spread. 

To remove this double counting effect, the charge per cluster can be defined in a different way. 
For each cluster, we build a summed waveform defined as the sum of the waveform amplitude seen in each pad of the cluster at all times.
We then take as a cluster charge estimator the maximum of this summed waveform. 
We refer to this method as WF$_{sum}$.
%Before: To remove the double counting, the charge per cluster can be defined in a different way. For each cluster, we build a summed waveform defined as the sum of the charge measured in each pad of the cluster at a given time and we take as a measurement of the charge per cluster the maximum of the summed waveform. We refer to this method as WF$_{sum}$.
%comment: a bit confusing to define a charge by using the word charge
%comment: I thought it was a all time

\begin{figure}[H]%!ht
    \centering
    \begin{minipage}{0.49\linewidth}
        \centering
        \includegraphics[width=\linewidth]{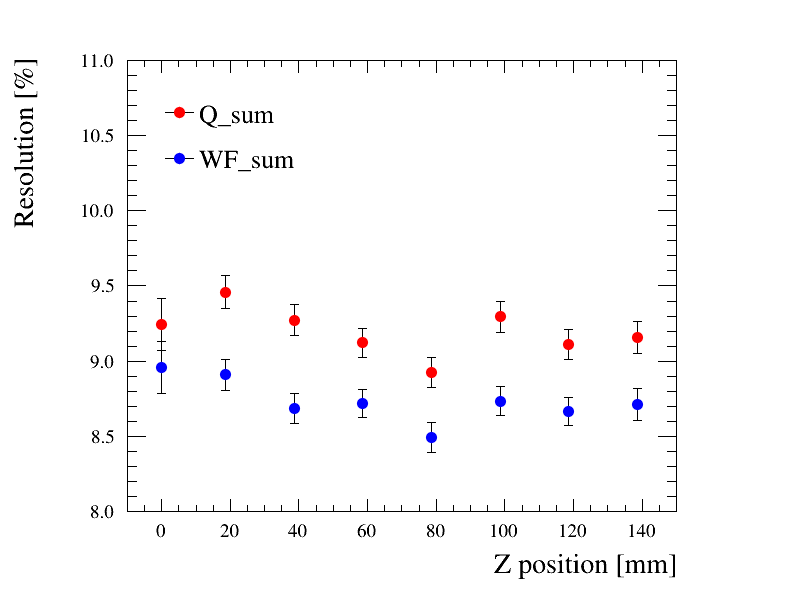} \\ a
    \end{minipage}
    \begin{minipage}{0.49\linewidth}
        \centering
        \includegraphics[width=\linewidth]{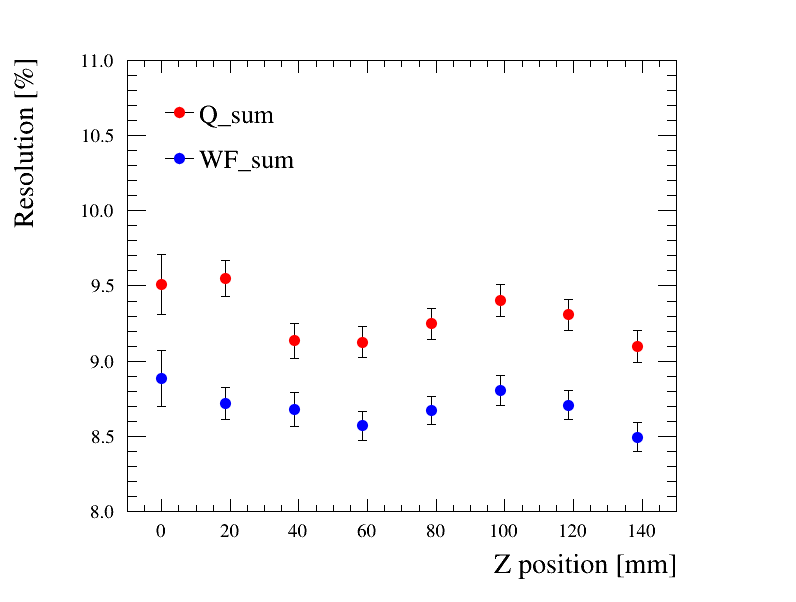} \\ b
    \end{minipage}
    \caption{
        dE/dx resolution with respect to the drift distance 
for the beam parallel to the pad side with a magnetic field of 0.2 T and peaking times of 200 ns (a) and 412 ns (b).
 Q$_{sum}$ method consists in summing the maximum of the waveform in each pad of the cluster while WF$_{sum}$ corresponds to maximum of the summed waveforms in a cluster. 
    }
    \label{fig:dEdx_driftD}
\end{figure}

A comparison of the results of the two methods as a function of the drift distance is shown in \autoref{fig:dEdx_driftD} for the two different peaking times and for data taken with 0.2 T magnetic field and with the beam parallel to the ERAM side. 
It can be seen that with both methods, the resolution is well below 10\% for all the drift distances and, as expected, the WF$_{sum}$ method gives a better resolution comprised between 8.5\% and 9.0\%.

%%%%%%%%%%%%%%

\subsection{Deposited energy resolution for inclined tracks}

As explained in \autoref{sec:spread}, in order to reconstruct inclined tracks, different clustering algorithms are used. 
In the case of the deposited energy resolution, the usage of such algorithms has two impacts: a larger number of clusters per track are reconstructed but the track will have different paths in different clusters. 

\begin{figure}[H] 	
\centering
\includegraphics[scale = 0.3]{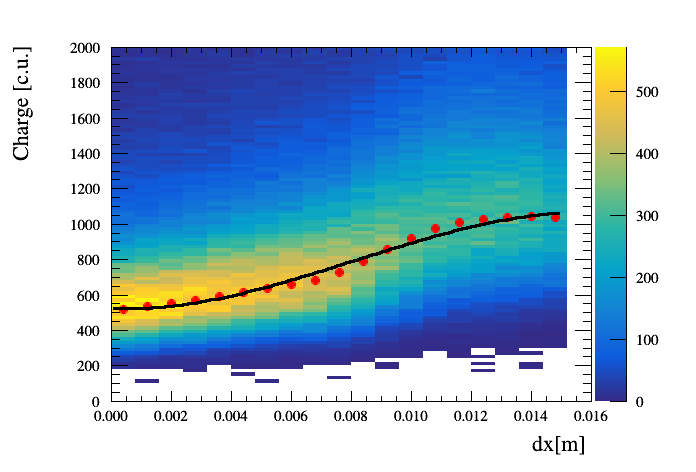}
\caption{The distribution of the charge with respect to the $dx$ of the track inclined by 45 degrees with respect to the pad plane}
\label{fig:dEdx_qvsdx}
\end{figure}

In the column or row clusters (defined in \autoref{fig:spatial:clusters}), tracks with the same angle with respect to the pad sides have the same $dx$ in each cluster (neglecting the curvature induced by the magnetic field).
%Before: In the clusters defined as the sum of pads in the same column (or row) in fact, tracks with same angle with respect to the pad plane, have the same $dX$ in each cluster (neglecting the curvature induced by the magnetic field). 
This is not true for diagonal clusters in which the $dx$ can vary between 0 and the diagonal of the pad ($\sim1.5~\textrm{cm}$). 
The distribution of the charge as a function of $dx$ for tracks inclined by 45 degrees with respect to the pad plane is shown in \autoref{fig:dEdx_qvsdx}. 
%previously: The distribution of the charge with respect to the dX for tracks inclined of 45 degrees with respect to the pad plane is shown in \autoref{fig:dEdx_qvsdx}. 
%other proposal: This figure reveals a clear dependence but it is not linear as one would expect from the simple consideration that the deposited energy should be proportional to the path. 
It is clear from this figure that there is a dependence, although this dependence is not linear as one would expect from the simple consideration that the deposited energy should be proportional to the path. 

The non-linearity is due to the fact that each cluster sees not only the direct charge due to the primary ionization, but it also sees some charge due to the spread on the resistive plane and to the transverse diffusion in the gas. 

In order to correct the $dx$ in each cluster, we fit the charge in each slice of $dx$ with a Landau function and we take the Most Probable Value (MPV). 
The distribution of MPV as a function of $dx$ is then parametrized with a third degree polynomial. 
In each cluster, the charge is corrected to take into account the real path length $dx$ and then the truncated mean is computed.

\begin{figure}[H]
    \centering
    \begin{minipage}{0.49\linewidth}
        \centering
        \includegraphics[width=\linewidth]{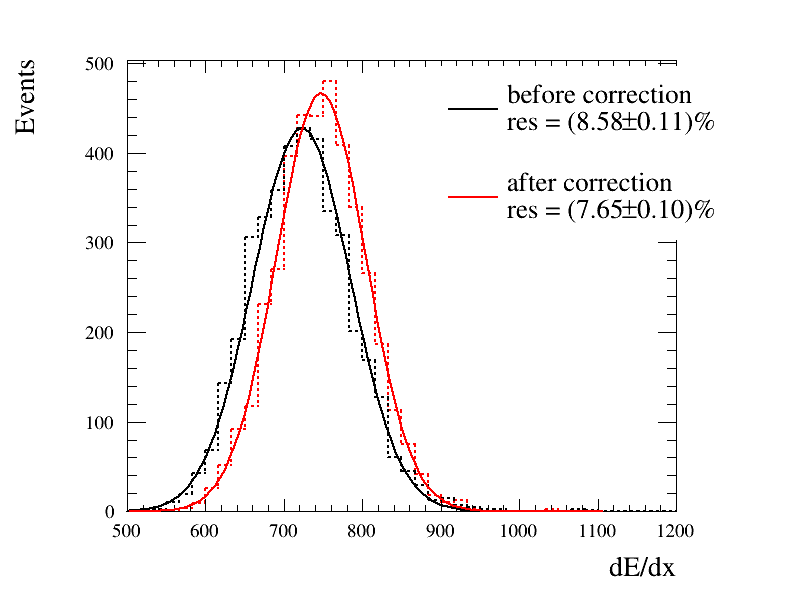} \\ a
    \end{minipage}
    \begin{minipage}{0.49\linewidth}
        \centering
        \includegraphics[width=\linewidth]{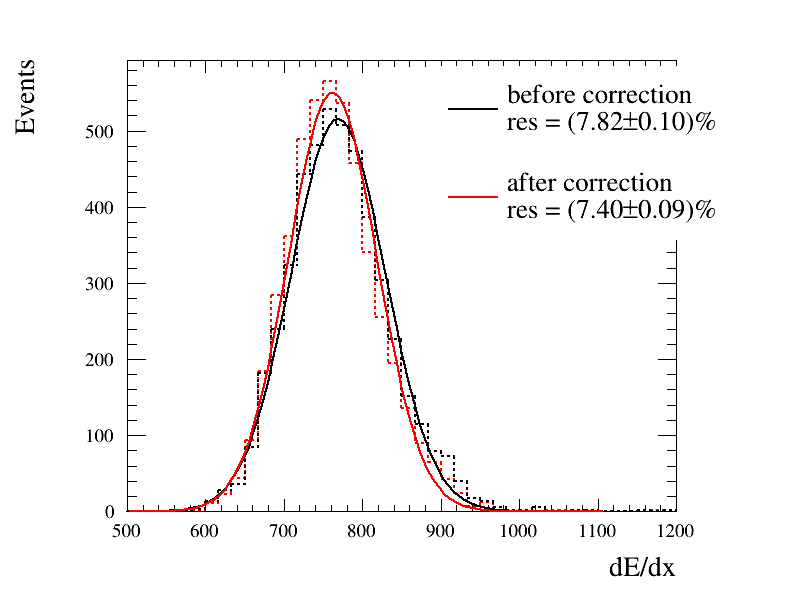} \\ b
    \end{minipage}
    \caption{The $dE/dx$ distribution for 40 deg inclined tracks at an electronics peaking time of 200 ns (a) and 412 ns (b) with and without the correction for $dx$. }
    \label{fig:dE_dxcorrected}
\end{figure}

The $dE/dx$ distributions for diagonal clustering with and without the $dx$ correction are shown in \autoref{fig:dE_dxcorrected}.

\begin{figure}[H]
    \centering
    \begin{minipage}{0.49\linewidth}
        \centering
        \includegraphics[width=\linewidth]{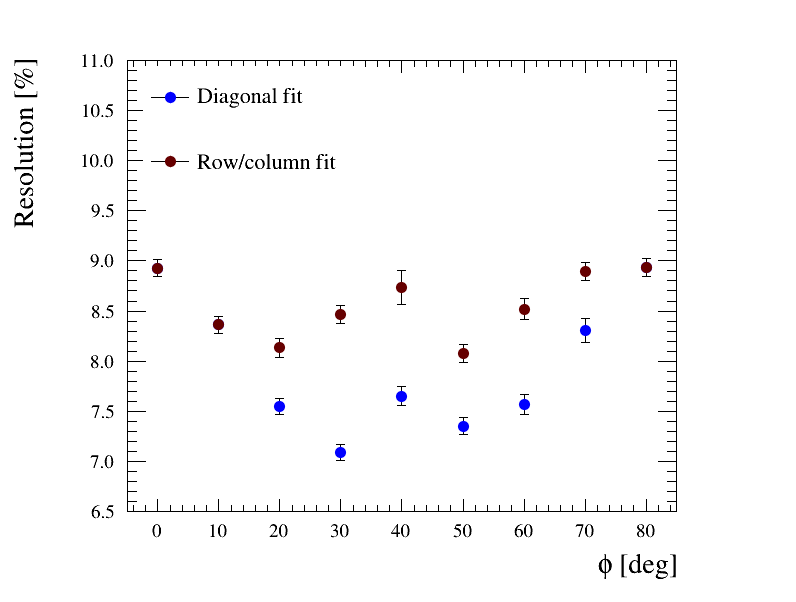} \\ a
    \end{minipage}
    \begin{minipage}{0.49\linewidth}
        \centering
        \includegraphics[width=\linewidth]{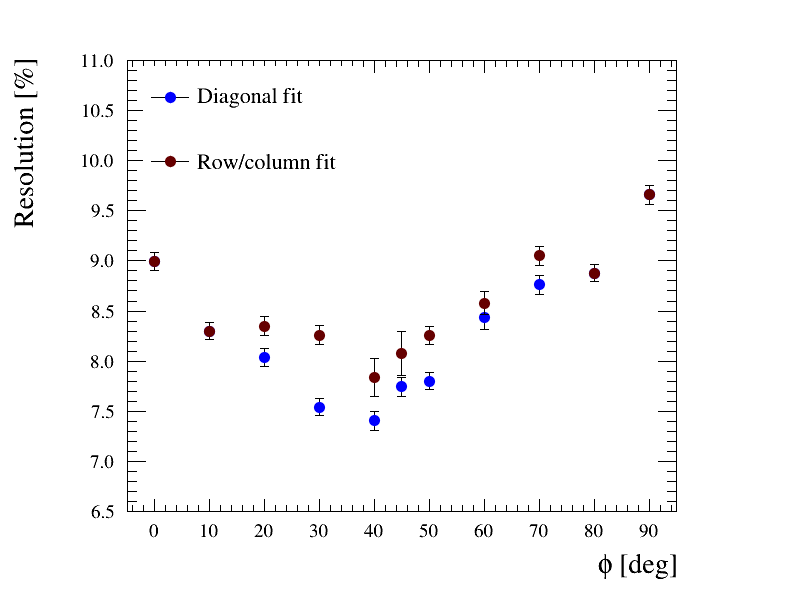} \\ b
    \end{minipage}
    %why say horizontal and inclined tracks??
    %need to define phi
    \caption{$dE/dx$ resolution versus the angle w.r.t the pad plane using column/row clustering or diagonal clustering.
    Column clustering is used from 0 to 40 deg and after 40 deg row clustering is used. 
    %Previously: dE/dx resolution versus the angle w.r.t pad plane using horizontal or diagonal clustering.
    Runs at a peaking time of 200 ns (a) and 412 ns (b), with a magnetic field of 0.2 T applied to the TPC prototype. 
    Diagonal clustering is corrected for the $dx$ as described in the text.}
    \label{Fig:dEdx_angle}
\end{figure}

The deposited energy resolution as a function of the beam inclination w.r.t. the pad plane is shown in \autoref{Fig:dEdx_angle}. 
As expected, the diagonal clustering, after proper $dx$ correction, provides the best resolution thanks to the larger amount of clusters in which the track is sampled.
%before: The deposited energy resolution as a function of the angle with respect to the pad plane is shown in \autoref{Fig:dEdx_angle}. As expected, the diagonal clustering, if properly corrected for the $dX$, provides the best resolution thanks to the larger amount of clusters in which the track is sampled.

\subsection{Dependence of the $dE/dx$ resolution on the number of clusters}

In this test beam, only one ERAM module was used. In the HA-TPC that will be installed at ND280, most of the tracks will cross two ERAM modules before exiting the TPC resulting in a larger number of clusters (72 for tracks parallel to the pad plane). The observed dependence of the deposited energy on the number of clusters can then be used to extrapolate the expected resolution in the HA-TPCs. 

\begin{figure}[H]
    \centering
    \begin{minipage}{0.49\linewidth}
        \centering
        \includegraphics[width=\linewidth]{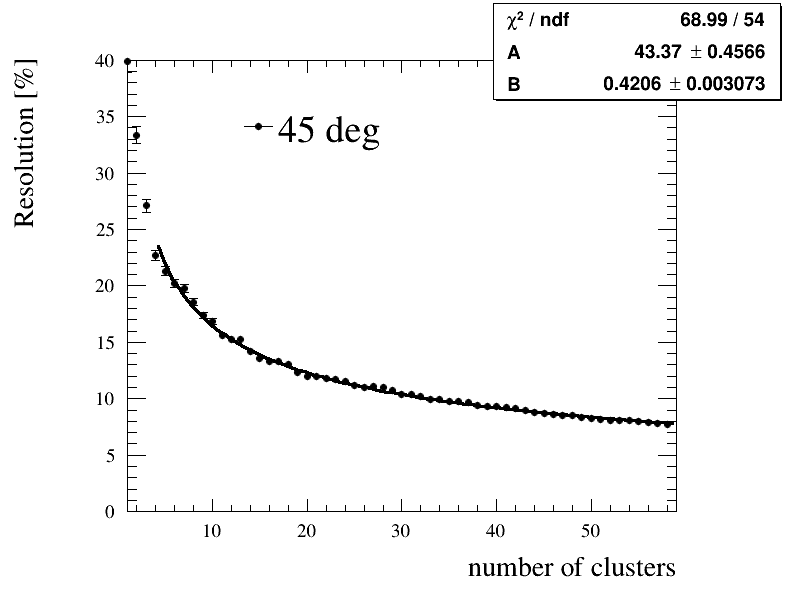} \\ a
    \end{minipage}
    \begin{minipage}{0.49\linewidth}
        \centering
        \includegraphics[width=\linewidth]{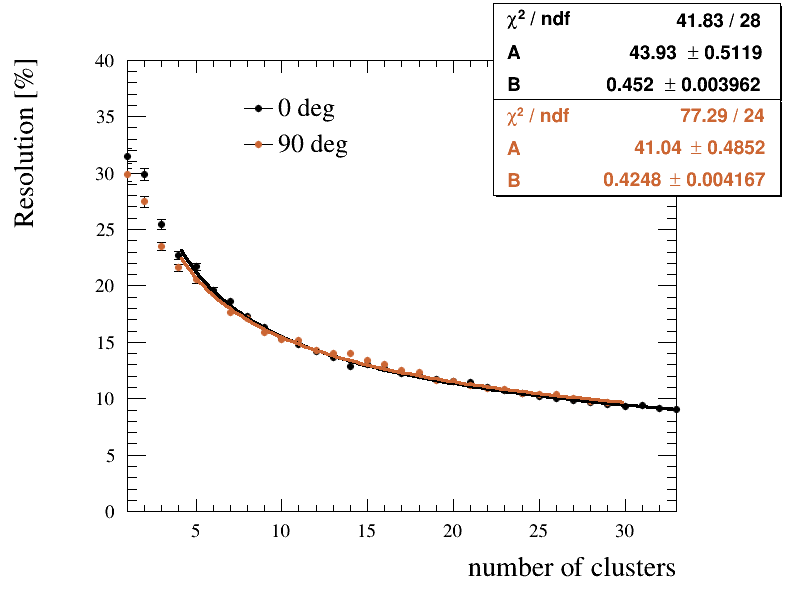} \\ b
    \end{minipage}
    %why say horizontal and inclined tracks??
    %need to define phi
    \caption{$dE/dx$ resolution versus the number of clusters for runs at a peaking time of 412 ns with a magnetic field of 0.2T applied to the TPC prototype. 
    Diagonal clustering (including $dx$ correction) is used for 45 deg inclined tracks (a).
     In (b), column clustering is employed for 0 deg horizontal tracks and row clustering for 90 deg vertical tracks (b). 
    %before: The diagonal clustering is used for 45 deg inclined tracks (a) and horizontal clustering is used for 0 deg horizontal tracks and 90 deg vertical tracks (b) are compared.  
    %Previously: The horizontal and diagonal clustering for 45 deg (a) and horizontal for 0 and 90 deg (b) are compared.  
    }
    \label{fig:dEdx_vs_ncluster}
\end{figure}

This can be done by computing the truncated mean using only a fraction of the available clusters. 
The dependence of the truncated mean on the number of clusters used for different samples is shown in \autoref{fig:dEdx_vs_ncluster}. 
%previously: The dependence of the truncated mean as a function of the number of clusters for different samples is shown in \autoref{fig:dEdx_vs_ncluster}. 

The resulting distribution of the deposited energy resolution as a function of the number of clusters $N$ is then fit with the function:
$f(N)=A N^{-B}$. $B$ equal to 0.5 would correspond to a simple $\sqrt{N}$ dependence.

In the case of horizontal or vertical tracks, we observe similar behavior: the horizontal tracks being slightly better because of the larger pad size in the horizontal direction.
Extrapolated to two ERAM modules, we obtain a deposited energy resolution of the order of 6\% for all the angles.
%other proposal: Extrapolated to two ERAM modules, the inclined tracks result leads to a deposited energy resolution of 5.42\% for diagonal clustering.
%before: Extrapolated to two ERAM modules for inclined tracks result in a deposited energy resolution of 5.42\% for diagonal clustering.
%comment: could be clearer 
%For inclined tracks, we can compare the performances of horizontal and diagonal clustering. Extrapolated to two ERAM modules, this result in a deposited energy resolution of XXX\% for horizontal clustering and XXX\% for diagonal clustering. 

%% file: rcmap.tex
% !TEX root = resistiveMM.tex

%Editor: Samira
%%%%%%%%%%%%%%%%%
%\subsection{RC map}
%%%%%%%%%%%%%%%%%
The quantity controlling the charge spreading over time is the product $RC$ (\autoref{sec:Micromegas}.  %where $R$ is the surface resistivity of the layer and $C$ the capacitance determined by the spacing between the anode and readout planes.   
To have a better understanding of our detector, we reconstruct the map of $RC$ using horizontal tracks in data where a scan in y-direction was performed at peaking time of 412 $\textrm{ns}$. 
This map is crucial to characterize our detector and its uniformity and is also needed for a detailed detector simulation.\\
To extract these $RC$ values, we use an analytical model of the charge dispersion, adjusted to the waveforms measured in the pads. 
The  induced  charge  on  a  rectangular  pad  below  the  resistive  layer can  be  calculated by integrating the charge density function over the pad area ~\cite{Dixit:2003qg} :
%%Problem of centering
%
%\begin{center}
\begin{equation}
\begin{split}
\mathcal{Q}(t)=\frac{q_{e}}{4}\left[erf( \frac{ x_\textup{high}-x_{0}}{\sqrt{2}\sigma(t)}) -erf( \frac{ x_\textup{low}-x_{0}}{\sqrt{2}\sigma(t)} )\right]\times \\ 
\left[erf( \frac{ y_\textup{high}-y_{0}}{\sqrt{2}\sigma(t)}) -erf( \frac{ y_\textup{low}-y_{0}}{\sqrt{2}\sigma(t)} )   \right]
\label{eq:chargedispersion}
\end{split}
\end{equation}
%\end{center}
%%%%%%%%%%%%
with $q_{e}$ is the initial charge, ($x_{0}$,$y_{0}$) the track position, $x_\textup{high}$, $x_\textup{low}$, $y_\textup{high}$,  $y_\textup{low}$ the pad boundaries. In the denominator $\sigma(t)=\sqrt{(2t/\tau)+ \omega^{2}}$, the term $\tau= RC$  where $R$ is the surface resistivity of the layer and $C$ the capacitance determined by the spacing between the anode and readout planes. Finally, $\omega$ is associated to the transverse diffusion term.  \\

To compare to data, the characteristics of the front-end charge preamplifiers need also to be included. Longitudinal diffusion increases the size of electron charge clusters in the drift direction.  The longitudinal diffusion is neglected here since we have only 15 $\textrm{cm}$  drift distance.
The parameterization of the electronics shaping time effects $\mathcal{I}(t)$ is obtained from the simulation. The convolution of $\mathcal{I}(t)$ and $\mathcal{Q}(t)$ results in the full theoretical model, which is compared to the data. This convolution is handled numerically.
The fit is based on clusters of pads perpendicular to the track. Each cluster consists of a so-called leading pad collecting essentially the initial charge deposit, and in so-called neighbour pads sensitive mostly to the induced charge due to the resistive effect, as shown in~\autoref{fig:DESY_waveform}.

\begin{figure}[!ht]
    \centering
    \begin{minipage}{0.32\linewidth}
        \centering
        \includegraphics[width=\linewidth]{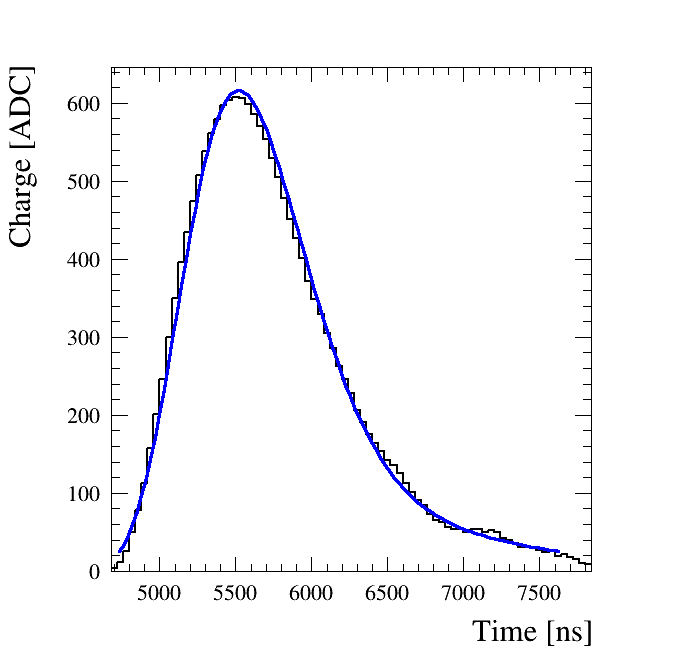} \\ a
    \end{minipage}
    \begin{minipage}{0.32\linewidth}
        \centering
        \includegraphics[width=\linewidth]{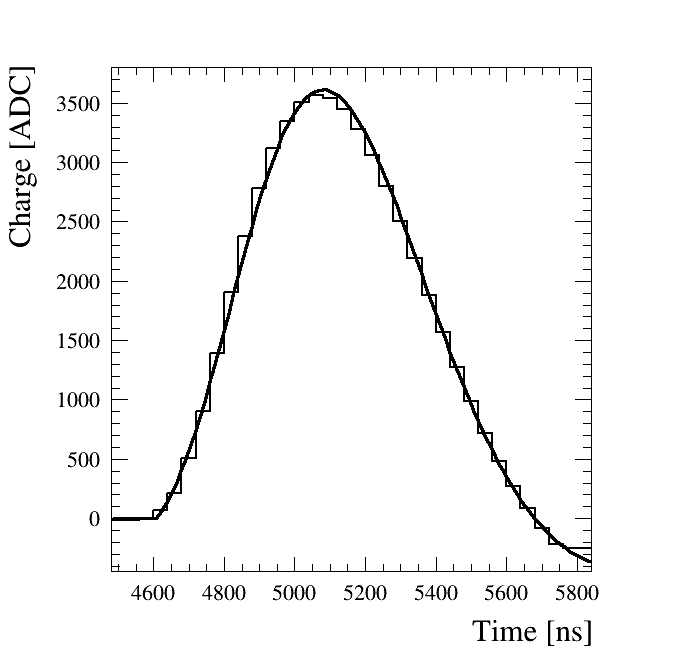} \\ b
    \end{minipage}
    \begin{minipage}{0.32\linewidth}
        \centering
        \includegraphics[width=\linewidth]{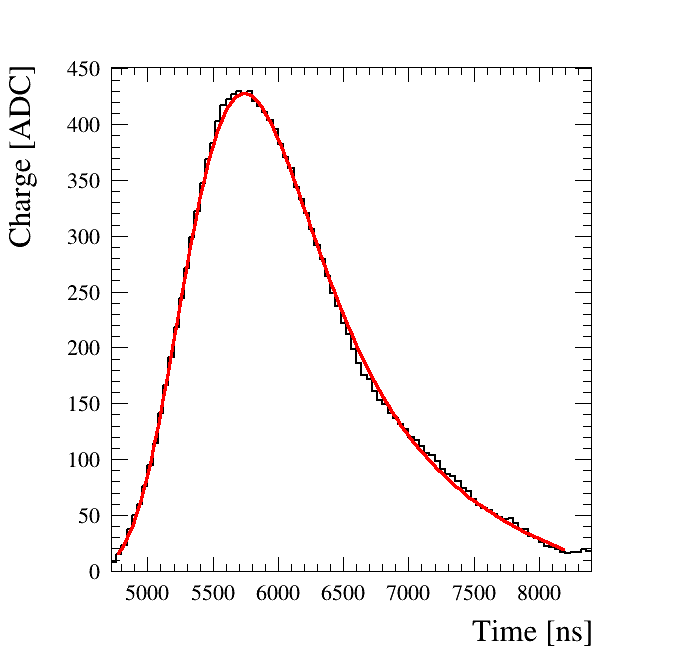} \\ c
    \end{minipage}
    
    \caption{Example of waveform fit results for the leading pad (b) and its neighbors (a) and (c) in a given cluster.}
    \label{fig:DESY_waveformfit}
\end{figure}

The fit procedure is as follows: we first fit the leading pad waveform with the electronics response function $\mathcal{I}(t)$, then we fit simultaneously the two neighbouring pads waveform with a convolution of $\mathcal{I}(t)$ with $\mathcal{Q}(t)$ in order to extract $RC$. Then a simultaneous fit of two neighbouring pads waveform use separate $\mathcal{Q}(t)$ functions, as the distance to the track can be different for the two pads, but we consider $RC$ and the electronics response parameters as common in the fit. 
Since we are using horizontal tracks, we can only fit the $y_{0}$-position of the track and we do not have any constraints on its $x$-position.
The track position $y_{0}$ is obtained with the PRF $\chi^2$ minimization method (see equation ~\eqref{eq::spatial:chi2}).
 \autoref{fig:DESY_waveformfit} shows an example of waveform fit results for the leading pad and its neighbours. 
%%%

Another method can be used to cross-check the $RC$ values obtained with the analytical model. This alternative method studies two parameters related to the signal propagation: the time at which the signal in neighbour pads is maximal and the ratio of amplitudes of the neighbour pads and the leading pad.
The time difference between the leading pad and one of its neighbour pad is found to be proportional to $RC$ as can be seen in the following formula:
$\Delta t_{1}-\Delta t_{2} = RC\times L \times y_{0}$, with $L$ the pad length and $y_{0}$ the track position and $\Delta t_{1,2}= t_\textup{Leading Pad}-t_\textup{Neighbor}$ the time difference between the leading pad and the neighbor pad.\\

\begin{figure}[!ht]
    \centering
    \begin{minipage}{0.49\linewidth}
        \centering
        \includegraphics[width=\linewidth]{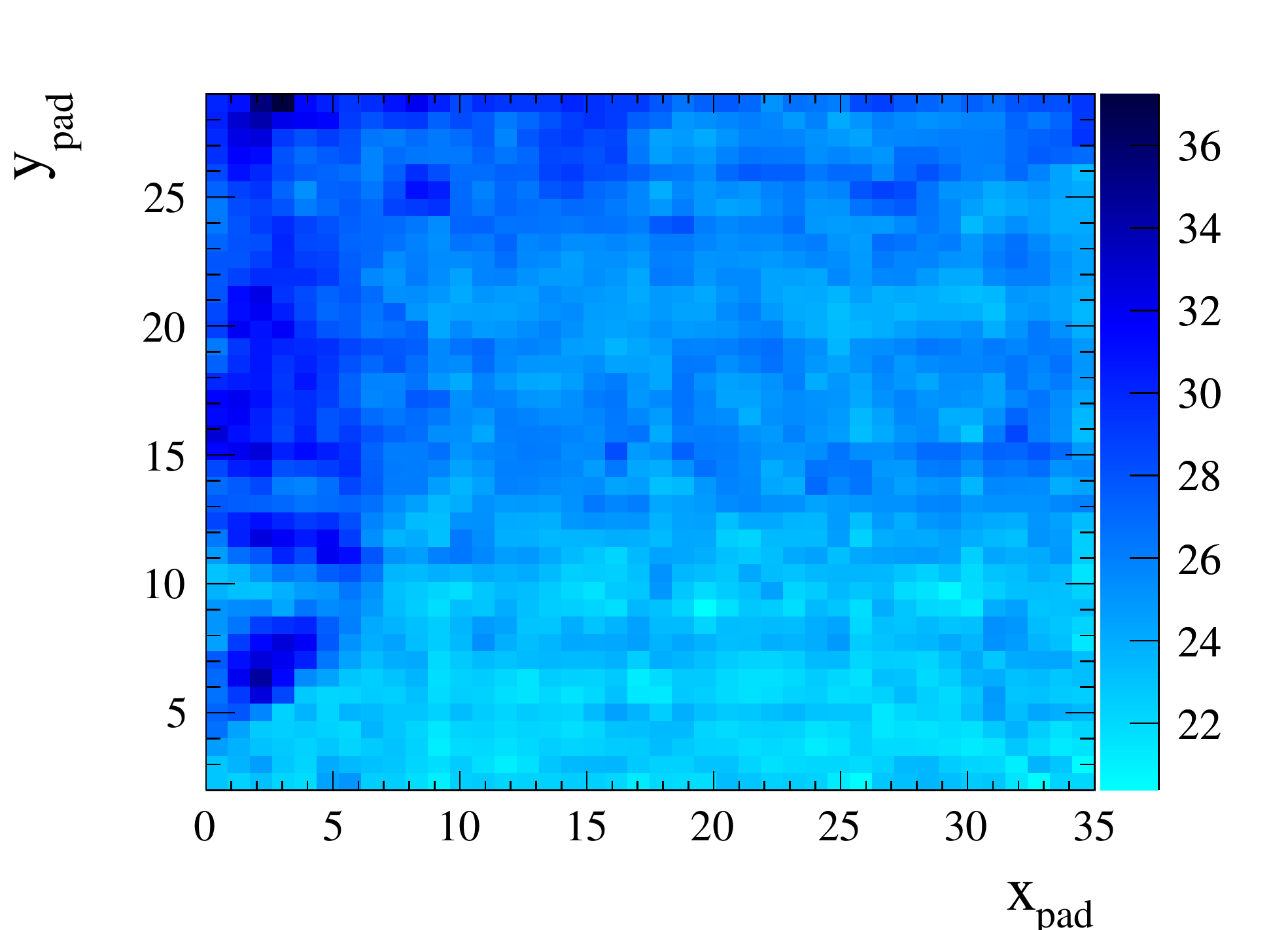} %%{RCmap_fit.png}
    \end{minipage}
    \begin{minipage}{0.49\linewidth}
        \centering
        \includegraphics[width=\linewidth]{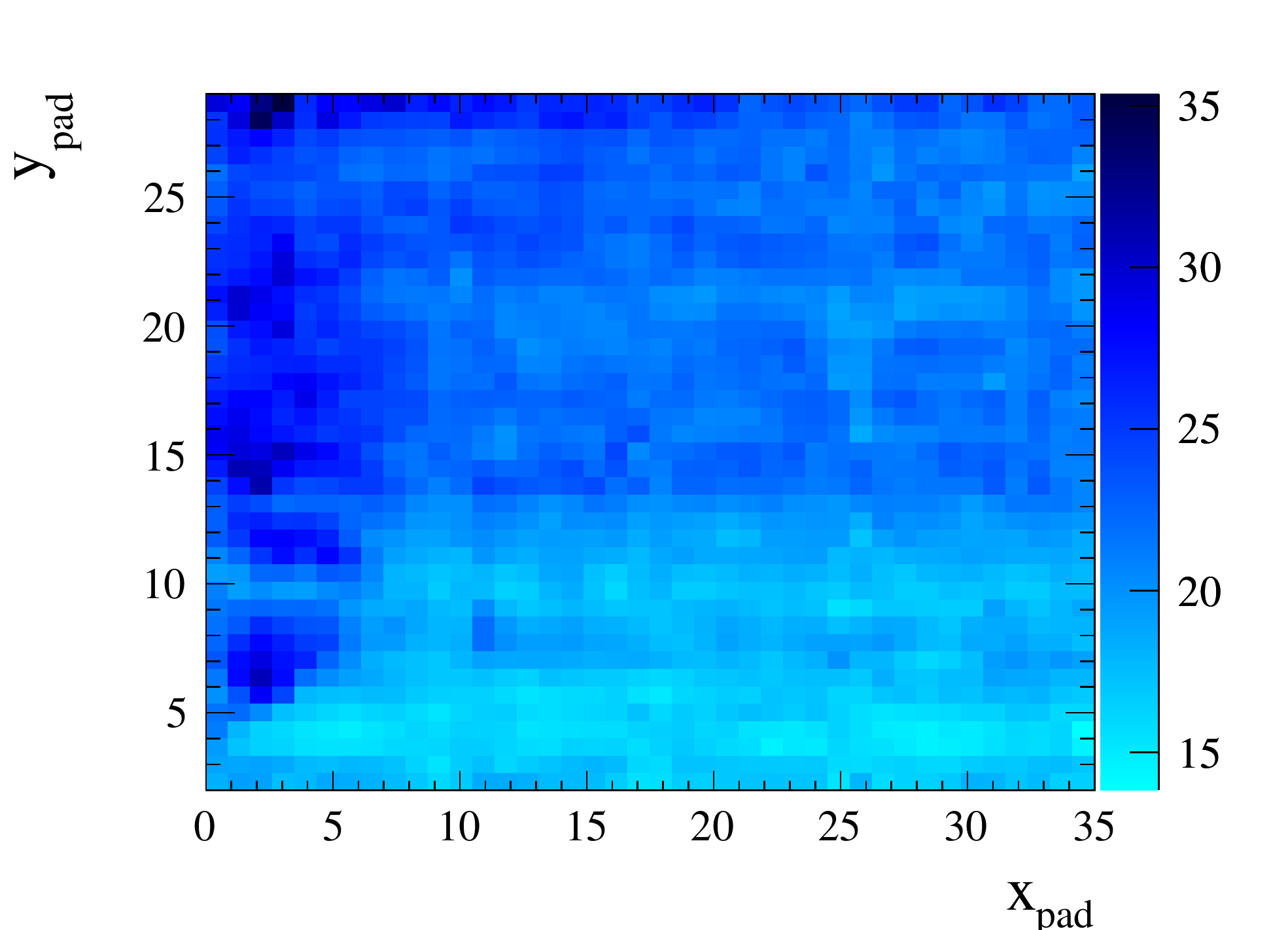} %%%{RCmap_2ndwayresultsRC.png}
    \end{minipage}
    \caption{The RC ($\textrm{ns/mm}^{2}$) map obtained using the fit from the analytical model (left) and the time difference method (right) described in the text.}
    \label{fig:DESY_RCmap}
\end{figure}

The $RC$ maps obtained using the analytical model and the time difference method described above are shown in~\autoref{fig:DESY_RCmap}. The $RC$ values are given in unit of $\textrm{ns}/\textrm{mm}^{2}$.
Both methods give results of the same order of magnitude though the second method is less precise than the fit procedure. A non-uniformity of $RC$ up to 30\% is observed using both methods.

\begin{figure}[!ht]
    \centering
    \includegraphics[width=\linewidth]{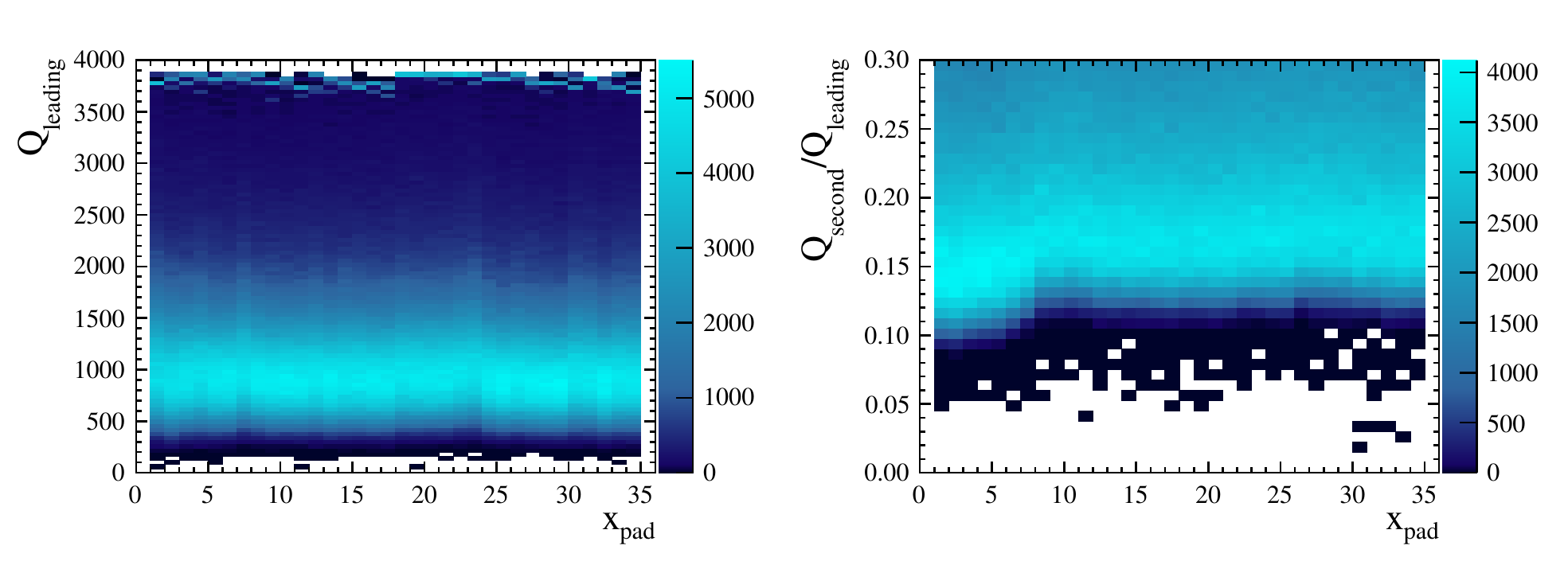}
    \caption{The distribution of the charge in the leading pad (left) and the charge fraction in the second to leading pad (right) versus the column for horizontal beam tracks.}
    \label{fig:charge_vs_column}
\end{figure}

This non-uniformity is confirmed with the observation of the charge collected in the second to leading pad. 
The charge observed in the leading pad is uniform across the detector while the measurements in the neighbours $Q_\textup{second}/Q_\textup{leading}$ demonstrate fluctuations, especially in the downstream detector region (\autoref{fig:charge_vs_column}). 
Lower charge fraction is consistent with the higher $RC$ value.

%% file: conclusions.tex
% !TEX root = resistiveMM.tex
We measured the performance of the ERAM prototype with beam particles at DESY. We studied both spatial and $dE/dx$ resolution as a function of the angle of the track with respect to the ERAM plane. We also characterized charge spreading and produced a RC map of the prototype.\\
Spatial resolution better than 600 $\text{\textmu}\textrm{m}$  is obtained for all the angles using a dedicated clustering algorithm which is adapted to the track angle.  Energy resolution better than 9\% is obtained for all the angles.  We expect ionization energy loss resolution to be better than 7\% for tracks crossing two ERAMs. 
Such performances fully satisfies the requirements for the upgrade of the ND280 TPC.